\documentclass[11pt,a4paper]{article}
\usepackage[centertags]{amsmath}
\usepackage{amsfonts}
\usepackage{amssymb}
\usepackage{amsthm}
\usepackage{graphicx}
\author{John Hammersley\footnote{J.C.Hammersley@dur.ac.uk} \\ \\ \textit{Department of Mathematical Sciences,} \\ \textit{Durham University,
South Road, Durham DH1 3LE UK}}
\title{\vspace{-1cm} \begin{flushright}
\footnotesize{hep-th/0609202} \\ \footnotesize{DCPT-06/25}
\end{flushright} \vspace{1cm} Extracting the bulk metric from boundary information
in asymptotically AdS spacetimes}
\date{}
\begin{document}

\maketitle

\begin{abstract}
We use geodesic probes to recover the entire bulk metric in certain
asymptotically AdS spacetimes. Given a spectrum of null geodesic
endpoints on the boundary, we describe two remarkably simple methods
for recovering the bulk information. After examining the issues
which affect their application in practice, we highlight a
significant advantage one has over the other from a computational
point of view, and give some illustrative examples. We go on to
consider spacetimes where the methods cannot be used to recover the
complete bulk metric, and demonstrate how much information can be
recovered in these cases.
\end{abstract}

\section{Introduction}

The holographic principle has inspired many ways of exploring
different spacetime geometries. The basic idea of holography, that
physics in a region of space can be described by the fundamental
degrees of freedom on its boundary, was originally applied to the
area of quantum gravity by 't Hooft \cite{thooft} and Susskind
\cite{suss}, but it was Maldacena \cite{adscft} who developed it
into something more tangible: the AdS/CFT correspondence
\cite{witten} \cite{polyakov} (For a comprehensive review, see
\cite{aha}).

Maldacena's conjecture postulated that string theory with $AdS_{5}
\times S^{5}$ boundary conditions is equivalent to $N = 4$ super
Yang-Mills theory, and AdS-type spacetimes have become the subject
of much work over the past decade. Many authors \cite[among
others]{Louko,kraus,levi,BHsing,kaplan,bala,hamilton,fest} have
examined how to extract information from behind the horizon of black
holes using the boundary correlation functions of the CFT. From the
spacetime point of view this involves using geodesics (both null and
spacelike) travelling from one boundary to another as probes.
However, this focus on singularities leaves open the question as to
how much of a non-singular spacetime can be recovered using geodesic
probes.\footnote{For more general work on extracting behind the
horizon information and holographic representations of objects in
AdS spacetimes see e.g.
\cite{banks,bala2,daniel,horo,freiv,hubeny,porr,maeda}.}

What we aim to explore in this paper is how accurately and
completely one can use the data contained within the boundary
correlators to recover the bulk metric in a static, spherically
symmetric, asymptotically AdS spacetime. We examine a variety of
different scenarios, and rather than merely looking at whether a
recovery of the metric is possible, we investigate the factors
affecting its recovery in practice.

We make use of the relation between null geodesic endpoints and
singularities in the correlation functions of operators inserted on
the boundary. In other words, that the correlation function between
any two points on the boundary which can be connected by a null
geodesic path diverges. Thus from the field theory we can identify,
say, the set of endpoints of null probes which all originated from
the same point on the boundary. This endpoint data can then be used
to recover information about the bulk metric.

In this paper we develop two methods for extracting this
information, and the essence of both is that one can focus on
specific radii by systematically using geodesics which penetrate to
different depths in the bulk. For null geodesics, there is only one
effective parameter which determines the minimum radius obtained by
the geodesic in a given spacetime, namely the ratio of the angular
momentum to the energy. When this ratio is close to one, the
geodesic probe travels around the edge of the spacetime; as it is
reduced, the probe travels deeper into the spacetime before
returning out to the boundary.

What we discover is that in certain types of metric, the process of
using null geodesic probes which penetrate to different depths can
be used to recover the entire bulk metric in a remarkably simple
manner.

We begin in section 2 by briefly reviewing some background material
on geodesics and their behaviour in a five dimensional AdS
spacetime. We observe that in this case, null geodesics always
travel between antipodal points on the boundary, as shown in
Fig.\ref{nullfig1}.

In section 3 we move on to considering modified spacetimes which are
still asymptotically AdS. In these metrics, null geodesics no longer
necessarily terminate at their antipodal points after travelling
through the bulk; we are presented with a spectrum of endpoints,
which provides the data we need to recover the metric. We discover
that on a plot of the endpoints (see Fig.\ref{modnullfig1} for an
example), the gradient at any point is related to the ratio of the
angular momentum and the energy of the corresponding null geodesic.
Using this relation together with the endpoint data, we devise a
simple iterative method for extracting the bulk metric, beginning
close to the boundary where the spacetime can be taken to be
approximately AdS. After considering whether this idea can also be
applied in a more general scenario, we go on to develop a second
method for iteratively recovering the metric.

These two methods are analysed in more detail in section 4, where we
examine a number of issues arising from using them in practice. We
propose several possible resolutions, in each case attempting to
strike a balance between the accuracy of the recovered metric
information and the computational effort required to do so. We then
describe a modification to the second method which greatly improves
the accuracy without a corresponding increase in computational
effort. Examples of the methods being used to recover the metric
information in two different spacetimes are then given in section 5,
where the superiority of the modified second method becomes apparent
(see figures \ref{examp1}-\ref{examp22} and the corresponding
tables).

Section 6 looks at limitations in the methods which arise not from
the numerics but from the geometry of certain spacetimes we would
like to consider. We examine how much of the bulk metric can be
recovered in setups where the metric gives a non-monotonic effective
potential for the geodesics, and where there is a singularity at the
centre of the spacetime.

In section 7 we discuss the results and mention possible extensions
of the project, such as using the endpoint data of spacelike
geodesics or considering non-static, non-spherically symmetric
spacetimes.

\section{Background}

The metric for a five-dimensional Anti-de Sitter spacetime is often
given in the form:

\begin{equation} \label{eq:AdSmetric}
ds^{2} = - f(r) dt^{2} + \frac{dr^{2}}{f(r)} + r^{2}d\Omega_{3}^{2}
\end{equation}
\begin{equation} \label{eq:f1}
f(r) = 1 + \frac{r^{2}}{R^{2}}
\end{equation}
where $R$ is the AdS radius.

To calculate the geodesic paths, we first suppress two of the
angular coordinates by setting the geodesics to lie in their
equatorial plane. Then from the form of the metric we can see the
existence of two Killing vectors, $\partial/\partial t$ and
$\partial/\partial \phi$, where $\phi$ is the remaining angular
coordinate, which lead to two constraints on the motion. The first
constraint is related to the energy and the second to the angular
momentum:

\begin{equation} \label{eq:AdSconstraint1}
E = f(r) \dot t
\end{equation}
\begin{equation} \label{eq:AdSconstraint2}
L = r^{2} \dot \phi
\end{equation}
where $\dot \space = \frac{d}{d\lambda}$, for some affine parameter
$\lambda$.

The geodesic paths can be found by extremizing the action

\begin{equation} \label{eq:action}
S = \int \, \dot x^{2} \mathrm{d} \lambda
\end{equation}
which leads to a third constraint on the motion, given by:

\begin{equation} \label{eq:AdSconstraint3}
\dot x^{2} = - f(r) \dot t^{2} + \frac{\dot r^{2}}{f(r)} + r^{2}\dot
\phi^{2}
\end{equation}
where $\dot x^{2} \equiv \kappa = +1,-1,0$ for spacelike, timelike
and null geodesics respectively.

By substituting \eqref{eq:AdSconstraint1} and
\eqref{eq:AdSconstraint2} into \eqref{eq:AdSconstraint3}, these
three equations can be combined together to eliminate $\dot t$ and
$\dot \phi$, giving:

\begin{equation} \label{eq:AdSveff1}
\kappa = - \frac{E^{2}}{f(r)} + \frac{\dot r^{2}}{f(r)} +
\frac{L^{2}}{r^{2}}
\end{equation}
which can be rewritten so as to introduce the concept of an
effective potential for the geodesics:

\begin{equation} \label{eq:AdSveff2}
\dot r^{2} + V_{eff} = 0
\end{equation}
with

\begin{figure}
\begin{center}
  \includegraphics[width=0.8\textwidth]{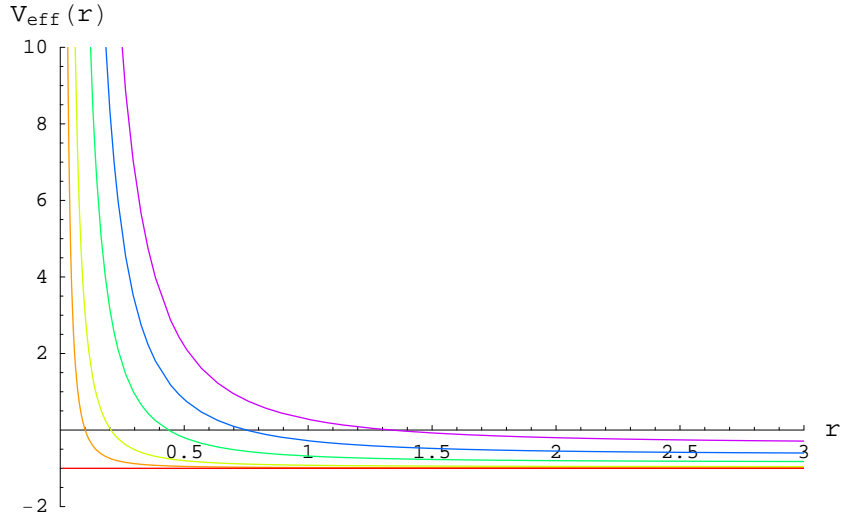}\\
\end{center}
\caption{Plot of the effective potential in pure AdS (with $R = 1$)
for a sample of null geodesics, with $E = 1.0$ and $L =
0.0,0.1,0.2,0.4,0.6$ and $0.8$}\label{Vefffig1}
\end{figure}

\begin{equation} \label{eq:AdSveff3}
V_{eff} = - \left(f(r) \kappa + E^{2} - \frac{f(r)
L^{2}}{r^{2}}\right)
\end{equation}

Note that the RHS of \eqref{eq:AdSveff2} is zero, and so any part of
the effective potential which is positive represents a potential
barrier, see Fig.\ref{Vefffig1}. The minimum value of r obtained by
spacelike and null geodesics incoming from $r = \infty$ is given by
the largest solution to $V_{eff} = 0$, and is the endpoint (starting
point) of the ingoing (outgoing) part of the geodesic. The paths of
geodesics can be calculated and plotted numerically using these
equations, and for null geodesics in pure AdS, they can also be
calculated analytically (see appendix A).

\begin{figure}
\begin{center}
  \includegraphics[width=0.8\textwidth]{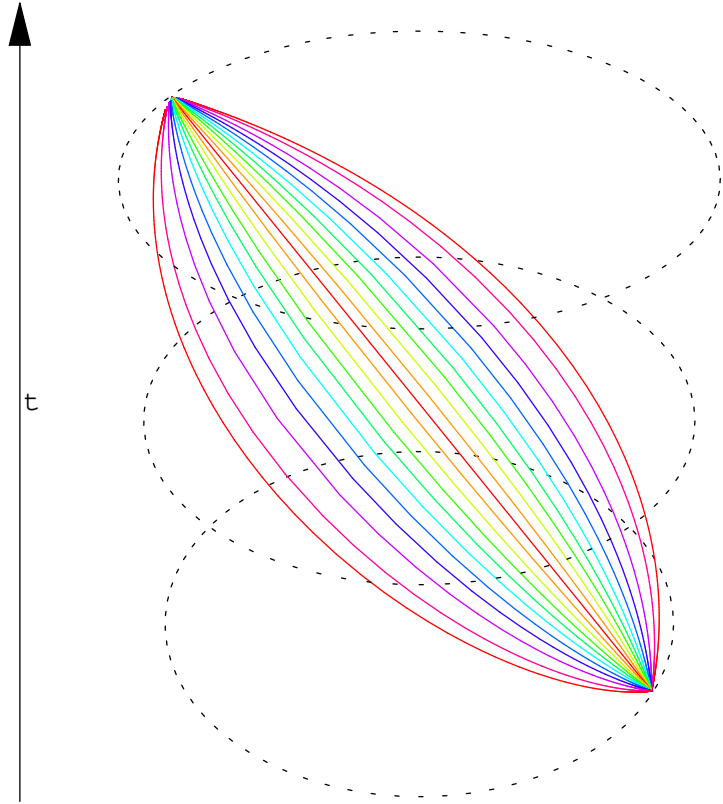}\\
\end{center}
\caption{A plot of some null geodesic paths passing through an AdS
spacetime, all starting from the same point $t=0$, $\phi = 0$ on the
boundary at $r = \infty$ (compactified to lie at $\arctan(\infty) =
\frac{\pi}{2}$ in the diagram).}\label{nullfig1}
\end{figure}

For null geodesics in AdS space, their endpoints always lie at the
antipodal point no matter what the ratio of angular momentum to the
energy. They can be plotted using the compactification given in
appendix A, which rescales the boundary at $r = \infty$ to lie on a
circle of radius $\pi/2$, see Fig.\ref{nullfig1}.

As mentioned in the introduction, the endpoints of null geodesics
correspond to singularities in the set of correlation functions in
the field theory lying on the boundary of the spacetime. For the
rest of the paper we assume that the endpoint data we use can be
generated in such a fashion from the field theory.

\section{Asymptotically AdS spacetimes}

We can consider a small modification to the pure AdS spacetime
described by \eqref{eq:AdSmetric} and \eqref{eq:f1}, by replacing
\eqref{eq:f1} with:

\begin{equation} \label{eq:f1new}
f(r) = 1 + \frac{r^{2}}{R^{2}}- p(r)
\end{equation}
where $p(r)$ is an analytic function which behaves as $r^{2}$ for
small $r$ and $r^{-2}$ for large $r$, such that the metric is
non-singular everywhere.\footnote{The $r^{2}$ behaviour of $p(r)$ at
small r ensures that we avoid a conical singularity at the origin.}

\subsection{Analysing the geodesic endpoints}

In this modified spacetime, null geodesic paths do not always end at
their antipodal point on the boundary; their final $t$ and $\phi$
coordinates depend on the ratio of the angular momentum ($L$) to the
energy ($E$). From a physical point of view this can be thought of
as the extra term $p(r)$ representing an attractive modification to
the centre of the spacetime, such that the geodesics follow a more
curved path through the bulk, and thus ``overshoot'' the antipodal
point. Their extra path length then also accounts for the increase
in the time taken to reach the boundary at infinity.

\begin{figure}
\begin{center}
  \includegraphics[width=0.45\textwidth]{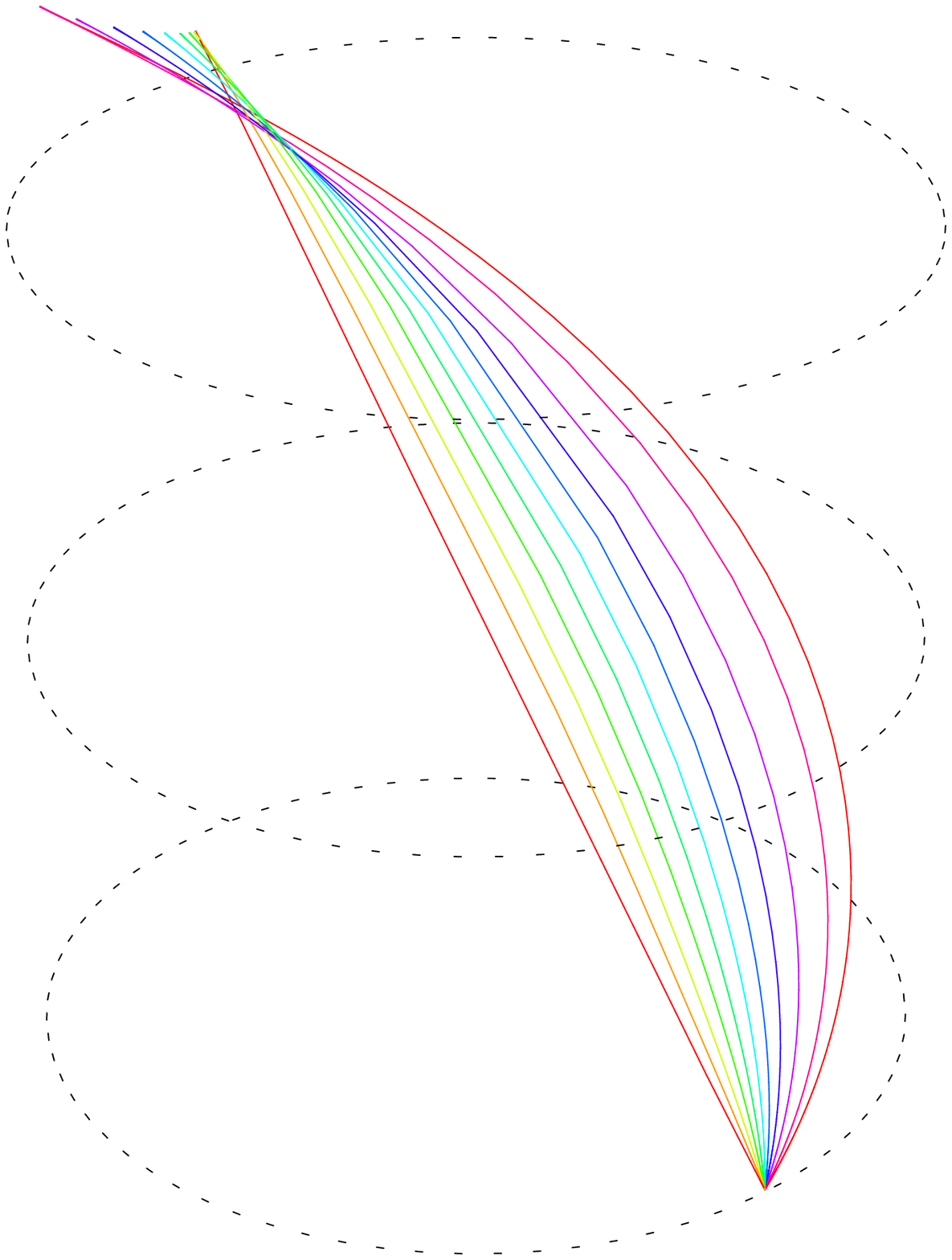}
  \includegraphics[width=0.45\textwidth]{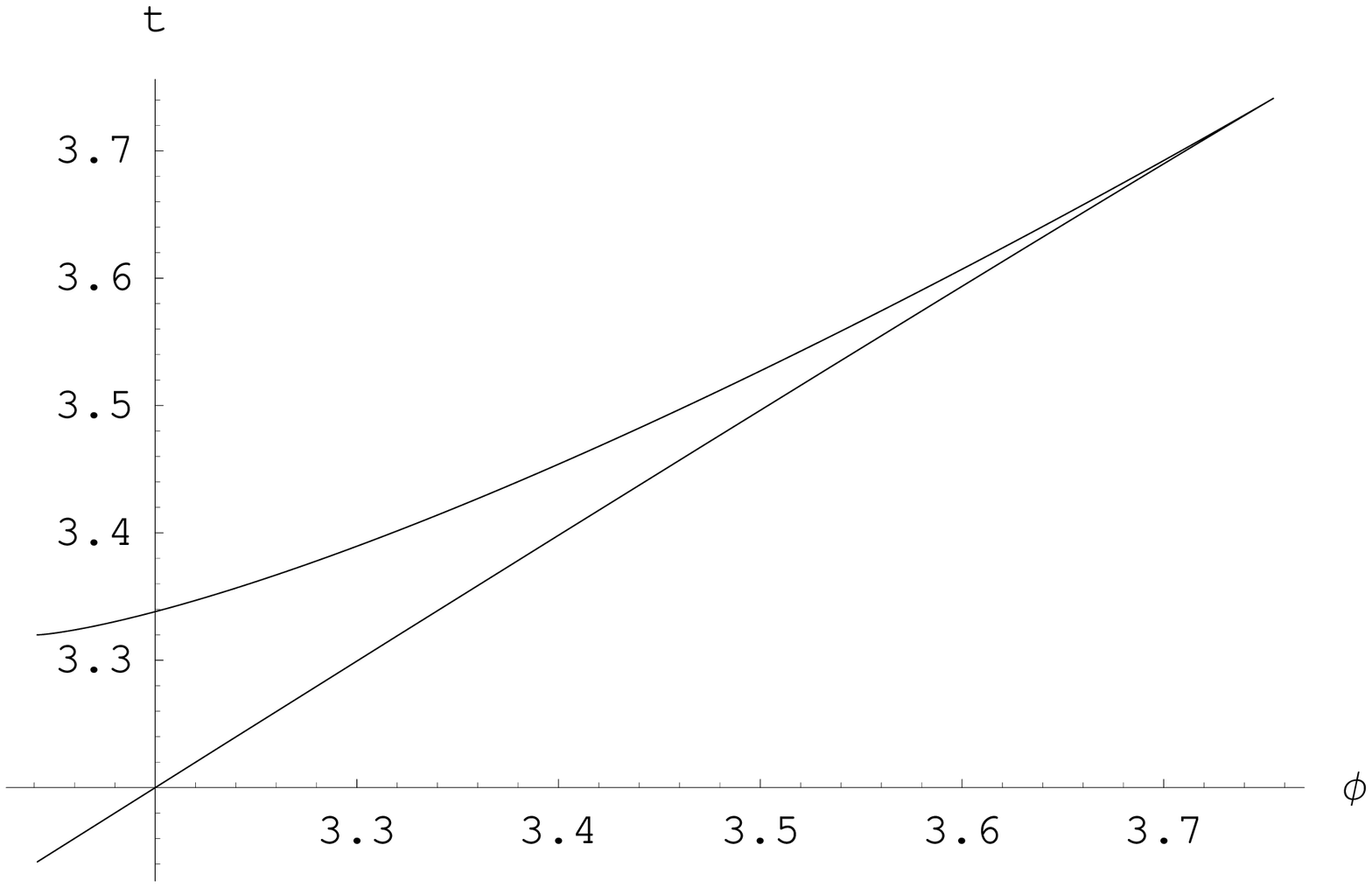}\\
\end{center}
\caption{A plot of the null geodesic paths passing through a
modified AdS spacetime, all starting from the arbitrary point $t=0$,
$\phi = 0$ on the boundary and with positive angular momentum. The
corresponding full spectrum of null geodesic endpoints for this
spacetime is shown on the right.}\label{modnullfig1}
\end{figure}

If we consider the set of null geodesics all beginning at the same
point, $(t_{0},\phi_{0})$ on the boundary, we can obtain the full
spread of endpoints by varying $L/E$ from 0 to 1.
Fig.\ref{modnullfig1} shows an example of the geodesic paths in a
modified AdS spacetime alongside a plot of their endpoints.

From this plot of the endpoints, what information about the bulk
metric can be recovered? For a geodesic beginning at
$(t_{0},\phi_{0})$ and ending at $(t_{1},\phi_{1})$ on the boundary,
we have:

\begin{equation} \label{eq:tr}
\int_{t_{0}}^{t_{1}}\, \mathrm{d} t = 2 \int_{r_{min}}^{\infty}
\frac{1}{f(r) \sqrt{1 - y^{2} \frac{f(r)}{r^{2}}}} \, \mathrm{d} r
\end{equation}
\begin{equation} \label{eq:phir}
\int_{\phi_{0}}^{\phi_{1}}\, \mathrm{d} \phi = 2
\int_{r_{min}}^{\infty} \frac{y}{r^{2} \sqrt{1 - y^{2}
\frac{f(r)}{r^{2}}}} \, \mathrm{d} r
\end{equation}
where $r_{min}$ is minimum radius obtained by the geodesic, and we
have introduced the parameter $y = L/E$ to denote the ratio of the
angular momentum to the energy.

We note that for a fixed metric, the final time and angular
coordinates will be functions of $y$ only, and so we can write:

\begin{equation} \label{eq:ty}
\int_{t_{0}}^{t_{1}}\, \mathrm{d} t = t_{1} - t_{0} \equiv t(y)
\end{equation}
\begin{equation} \label{eq:phiy}
\int_{\phi_{0}}^{\phi_{1}}\, \mathrm{d} \phi = \phi_{1} - \phi_{0}
\equiv \phi(y)
\end{equation}

If we define the function $g(y,r)$ as:

\begin{equation} \label{eq:brevity}
g(y,r) = \frac{1}{\sqrt{1 - y^{2} \frac{f(r)}{r^{2}}}}
\end{equation}
and consider the derivative of \eqref{eq:tr} and \eqref{eq:phir}
with respect to $y$ we have:

\begin{eqnarray} \label{eq:dtr}
\frac{d t(y)}{dy}&& \hspace{-0.5cm} = 2 \frac{d}{dy}
\left(\int_{r_{min}}^{\infty} \frac{g(y,r)}{f(r)} \, \mathrm{d} r
\right)
\\
&& \hspace{-0.5cm} = 2 y \int_{r_{min}}^{\infty}
\frac{\left(g(y,r)\right)^{3}}{r^{2}} \, \mathrm{d} r -
\left(\frac{2 \, g(y,r)}{f(r)}\right)\Big|_{r = r_{min}} \frac{d
r_{min}}{dy}
\end{eqnarray}
and
\begin{eqnarray} \label{eq:dphir}
\frac{d \phi(y)}{dy} && \hspace{-0.5cm} = 2 \frac{d}{dy}
\left(\int_{r_{min}}^{\infty} \frac{y \, g(y,r)}{r^{2}} \,
\mathrm{d} r \right)
\\
&& \hspace{-0.5cm} = 2 \int_{r_{min}}^{\infty}
\frac{\left(g(y,r)\right)^{3}}{r^{2}} \, \mathrm{d} r -
\left(\frac{2 \, y \, g(y,r)}{r^{2}}\right)\Big|_{r = r_{min}}
\frac{d r_{min}}{dy}
\end{eqnarray}

Comparing the integral terms in the two equations above, we see that
they are identical upto a factor of $y$. We can then use the fact
that at $r = r_{min}$, $y^{2} = \frac{f(r_{min})}{r_{min}^{2}}$ to
rewrite the first equation as:

\begin{equation} \label{eq:dtrb}
\frac{d t(y)}{dy} = 2 y \int_{r_{min}}^{\infty}
\frac{\left(g(y,r)\right)^{3}}{r^{2}} \, \mathrm{d} r - y
\left(\frac{2 \, y \, g(y,r)}{r^{2}}\right)\Big|_{r = r_{min}}
\frac{d r_{min}}{dy}
\end{equation}

After checking that the divergent piece of the integral cancels with
the divergent second term in \eqref{eq:dtrb} (see appendix B) we
therefore have that:

\begin{equation} \label{eq:dtdphi}
\frac{d t(y)}{dy} = y \, \frac{d \phi(y)}{dy}
\end{equation}
which can be rewritten as

\begin{equation} \label{eq:dtdphib}
\frac{d t}{d \phi} = y(\phi)
\end{equation}

Thus from the plot of the endpoints, by taking the gradient at each
point we are able to obtain the value of $y$ for that geodesic. Note
that at no point in this derivation have we used the fact that the
spacetime is asymptotically AdS, and it can easily be shown that the
result holds for any static, spherically symmetric spacetime (i.e.
any metric of the form of \eqref{eq:newAdSmetric}).\footnote{It is
also not necessary for the boundary on which the geodesics begin and
end to be at infinity; the result is valid for any spherical
boundary within the bulk.} Calculations along these lines are also
not restricted to null geodesics, see \cite{fest} for similar
derivations involving spacelike ones. In the next section we will
see how we can further use this to recreate the metric function
$f(r)$.

\subsection{Reconstructing $f(r)$: Method I} \label{sec:method1}

The method involves iteratively recovering $f(r)$ starting from
large radius and working down towards $r = 0$. We are working in an
asymptotically AdS spacetime. This means that as $r \rightarrow
\infty$, $f(r) \rightarrow r^{2}/R^{2} + 1$; thus we can say that
for $r \ge r_{n}$ for some $r_{n}$ which can be arbitrarily large,
$f(r) \approx r^{2}/R^{2} + 1$. We set the AdS radius ($R$) to one,
and by considering large enough $y = y_{n}$ such that the minimum
radius, $r_{min}$, corresponds to $r_{n}$, \eqref{eq:tr} becomes
(setting the initial time, $t_{0}$, to be zero):

\begin{eqnarray} \label{eq:trm1}
t_{n} = \int_{0}^{t_{n}}\, \mathrm{d} t && \hspace{-0.5cm} = 2
\int_{r_{n}}^{\infty} \frac{g(y_{n},r)}{f(r)} \, \mathrm{d} r
\\ \label{eq:trm12}
&& \hspace{-0.5cm} \approx 2 \int_{r_{n}}^{\infty} \frac{1}{(r^{2} +
1) \sqrt{1 - y_{n}^{2} \frac{r^{2} + 1}{r^{2}}}} \, \mathrm{d} r =
\pi
\end{eqnarray}
as we would expect, as geodesics remaining far from the centre of
the space would not ``see'' the modification and thus behave as in
pure AdS. In this case we can solve the equation for the minimum
radius:

\begin{equation} \label{eq:gradm1}
y_{n}^{2} = \frac{r_{n}^{2}}{f(r_{n})} = \frac{r_{n}^{2}}{r_{n}^{2}
+ 1}
\end{equation}
to get $r_{n}$ in terms of $y_{n}$, which is determined from the
geodesic endpoints. Thus we can determine $r_{n}$ (and hence
$f(r_{n}) = r_{n}^{2} + 1$).

If we now consider a geodesic with slightly lower ratio of angular
momentum to energy, say with $y_{n - 1} = y_{n} - \epsilon$, where
$\epsilon > 0$, we have the equation:

\begin{equation} \label{eq:trnminus1m1}
t_{n-1} = 2 \int_{r_{n-1}}^{\infty} \frac{g(y_{n-1},r)}{f(r)} \,
\mathrm{d} r
\end{equation}
which can be split up as follows (noting that $r_{n-1} < r_{n}$):
\begin{equation} \label{eq:trnminus1bm1}
t_{n-1} = 2 \int_{r_{n-1}}^{r_{n}} \frac{g(y_{n-1},r)}{f(r)} \,
\mathrm{d} r + 2 \int_{r_{n}}^{\infty} \frac{g(y_{n-1},r)}{f(r)} \,
\mathrm{d} r
\end{equation}

The second integral can be evaluated by setting $f(r) = r^{2} + 1$
as in \eqref{eq:trm1}, and we have

\begin{equation} \label{eq:trnminus1ccm1}
2 \int_{r_{n}}^{\infty} \frac{g(y_{n-1},r)}{f(r)} \, \mathrm{d} r =
\pi - 2 \arctan{\left( \sqrt{\left(1 - y_{n-1}^{2}\right)r_{n}^{2} -
y_{n-1}^{2}}\, \right)}
\end{equation}

We can approximate the first integral by taking a Laurent expansion
about the point $r_{n-1}$. This gives, to lowest order:

\begin{eqnarray} \label{eq:trnminus1dm1}
\int_{r_{n-1}}^{r_{n}} \frac{g(y_{n-1},r)}{f(r)} \, \mathrm{d} r &&
\hspace{-0.5cm} \approx \int_{r_{n-1}}^{r_{n}} \frac{(r -
r_{n-1})^{-1/2}}{f(r_{n-1}) \sqrt{\frac{2}{r_{n-1}} -
\frac{f'(r_{n-1})}{f(r_{n-1})}}} \, \mathrm{d} r
\\
&& \hspace{-0.5cm} = \frac{2 \sqrt{r_{n} - r_{n-1}}}{f(r_{n-1})
\sqrt{\frac{2}{r_{n-1}} - \frac{f'(r_{n-1})}{f(r_{n-1})}}}
\end{eqnarray}
where we have used that $y_{n-1}^{2} = r_{n-1}^{2}/f(r_{n-1})$ and
integrated. This can be simplified further by writing\footnote{Note
that this linear approximation of the gradient is only fine whilst
$r_{n} - r_{n-1}$ is kept small.}:

\begin{equation} \label{eq:fgradm1}
f'(r_{n-1}) \approx \frac{f(r_{n}) - f(r_{n-1})}{r_{n} - r_{n-1}}
\end{equation}

Thus we have:
\begin{equation} \label{eq:trnminus1finalm1}
t_{n-1} \approx \frac{4 (r_{n} - r_{n-1})}{f(r_{n-1}) \sqrt{\frac{2
r_{n}}{r_{n-1}} - \frac{f(r_{n})}{f(r_{n-1})} - 1}} + \pi - 2
\arctan{\left( \sqrt{\left(1 - y_{n-1}^{2}\right)r_{n}^{2} -
y_{n-1}^{2}}\, \right)}
\end{equation}
which can be used in conjunction with $y_{n-1}^{2} =
r_{n-1}^{2}/f(r_{n-1})$ to calculate $r_{n-1}$ and $f(r_{n-1})$, as
we know $t_{n-1}$ and $y_{n-1}$ from the geodesic endpoints, and
$r_{n}$ and $f(r_{n})$ have already been calculated.

\begin{figure}
\begin{center}
  \includegraphics[width=0.8\textwidth]{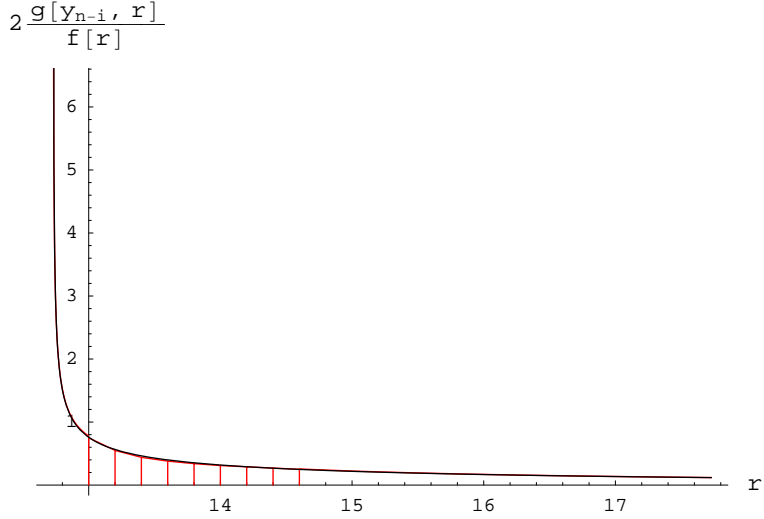}\\
\end{center}
\caption{A plot showing how the integrand of \eqref{eq:trnminusim1},
is split up into two end curves and a number of trapeziums in order
for the integral to be well approximated. The actual curve is shown
in black, with the approximations in red.}\label{traps1}
\end{figure}

For general $t_{n-i}$ we split the integral into several pieces; the
two ``end'' integrals, which we evaluate as in
\eqref{eq:trnminus1ccm1} and \eqref{eq:trnminus1dm1}, and a series
of integrals in the middle which can be evaluated using the
trapezium rule or similar (see Fig.\ref{traps1}). The formula for
general $t_{n-i}$ is then given by (for $i \geq 2$):
\begin{eqnarray} \label{eq:trnminusim1}
t_{n-i} && \hspace{-0.5cm} = 2 \int_{r_{n-i}}^{\infty}
\frac{g(y_{n-i},r)}{f(r)} \, \mathrm{d} r
\\
&& \hspace{-0.5cm} \approx A_{n-i} + B_{n-i} + C_{n-i}
\end{eqnarray}
where
\begin{equation} \label{eq:anminusi}
A_{n-i} = \frac{4 \, (r_{n-i+1} - r_{n-i})}{f(r_{n-i}) \sqrt{\frac{2
r_{n-i+1}}{r_{n-i}} - \frac{f(r_{n-i+1})}{f(r_{n-i})} - 1}}
\end{equation}
\begin{equation} \label{eq:bnminusi}
B_{n-i} = \sum_{j=1}^{i-1} \left(r_{n-j+1} - r_{n-j}\right) \left(
\frac{g(y_{n-i},r_{n-j+1})}{f(r_{n-j+1})} +
\frac{g(y_{n-i},r_{n-j})}{f(r_{n-j})} \right)
\end{equation}
\begin{equation} \label{eq:cnminusi}
C_{n-i} = \pi - 2 \arctan{\left( \sqrt{\left(1 -
y_{n-i}^{2}\right)r_{n}^{2} - y_{n-i}^{2}}\, \right)}
\end{equation}

For a spacetime of the form of \eqref{eq:f1new} with monotonic
effective potential, we can use the above to recover the function
$f(r)$ down to $r = 0$. See section 5 for examples.

\subsection{More general spacetimes}

So far we have restricted ourselves to considering metrics of the
form of \eqref{eq:AdSmetric}, however, we can look to extend this
further by considering the most general static, spherically
symmetric spacetimes, given by metrics of the form:

\begin{equation} \label{eq:newAdSmetric}
ds^{2} = - f(r) dt^{2} + h(r) dr^{2} + r^{2}d\Omega_{3}^{2}
\end{equation}
where the $dr^{2}$ coefficient is allowed to be different to the
$dt^{2}$ coefficient. Equations \eqref{eq:AdSconstraint1} and
\eqref{eq:AdSconstraint2} for the geodesics remain unchanged, and
the third is now given by:

\begin{equation} \label{eq:AdSveffc2}
\dot r^{2} - \left(\frac{\kappa}{h(r)} + \frac{E^{2}}{f(r) h(r)} -
\frac{L^{2}}{h(r) r^{2}}\right) = 0
\end{equation}
where $\kappa = 0$ for the null case. We can combine the equations
together as before to give the integral equations:

\begin{equation} \label{eq:trc}
\int_{t_{0}}^{t_{1}}\, \mathrm{d} t = 2 \int_{r_{0}}^{\infty}
\frac{1}{\frac{f(r)}{\sqrt{h(r)}} \sqrt{\frac{1}{f(r)} -
\frac{y^{2}}{r^{2}}}} \, \mathrm{d} r
\end{equation}
\begin{equation} \label{eq:phirc}
\int_{\phi_{0}}^{\phi_{1}}\, \mathrm{d} \phi = 2
\int_{r_{0}}^{\infty} \frac{y}{\frac{r^{2}}{\sqrt{h(r)}}
\sqrt{\frac{1}{f(r)} - \frac{y^{2}}{r^{2}}}} \, \mathrm{d} r
\end{equation}

The question now is whether we can use the two equations above along
with the expression $y^{2} = r_{0}^{2}/f(r_{0})$ relating the
parameter $y$ to the minimum radius to recover both $f(r)$ and
$h(r)$ and thus reconstruct the bulk metric from the endpoints. This
initially appears possible as we haven't explicitly used our
expression for $\phi$, \eqref{eq:phirc}, in our method of the
previous section. However, our observation that the gradient of the
endpoints is given by the angular momentum \eqref{eq:dtdphib}
suggests that we don't have three linearly independent relations.
Attempting to continue as before, by approximating the spacetime as
pure AdS at some large radius and working inwards, the degeneracy
soon becomes apparent (see appendix C), and we are unable to recover
bulk metric information in these more general spacetimes (unless we
\textit{a priori} know the relationship between $h(r)$ and $f(r)$).

\subsection{Reconstructing $f(r)$: Method II}
\label{sec:method2}

There is an alternative method for reconstructing $f(r)$ in a
spacetime of the form of \eqref{eq:AdSmetric} which makes the
degeneracy of the expressions for $t$ and $\phi$ more apparent. If
we start from expression \eqref{eq:dtdphi} and integrate over $y$ we
get:

\begin{equation} \label{eq:tphi}
\int^{t} \, \mathrm{d} t' = \int y \frac{d \phi}{d y} \, \mathrm{d}
y
\end{equation}
which can be integrated by parts:
\begin{equation} \label{eq:tphi2}
t(y) = y \, \phi(y) - \int \phi \, \mathrm{d} y
\end{equation}
and then rewritten using the expression for $\phi$ from
\eqref{eq:phir}:

\begin{equation} \label{eq:tphi3}
t(y) = y \, \phi(y) - \int \int_{r_{min}}^{\infty} \frac{2 y}{r^{2}
\sqrt{1 - y^{2} \frac{f(r)}{r^{2}}}} \, \mathrm{d} r \, \mathrm{d} y
\end{equation}

We can now reverse the order of integration, and as the function
$f(r)$ has no dependence on $y$, integrate over $y$:

\begin{equation} \label{eq:tphi4}
t(y) = y \, \phi(y) + \int_{r_{min}}^{\infty} \frac{2}{f(r)} \sqrt{1
- y^{2} \frac{f(r)}{r^{2}}} \, \mathrm{d} r
\end{equation}

Taking the initial conditions to be $(\phi_{0}, t_{0}) = (0,0)$, we
can say that for any endpoint $(\phi_{j}, t_{j})$ on the boundary,

\begin{equation} \label{eq:tphi5}
t_{j} - \frac{d t}{d \phi} \Big|_{(\phi_{j},t_{j})} \phi_{j} =
\int_{r_{j}}^{\infty} \frac{2}{f(r) \, g(y_{j},r)} \, \mathrm{d} r
\end{equation}
where we have renamed the minimum $r$ as $r_{j}$ and used our
definition for $g(y,r)$ from before. This, coupled with the equation
for the minimum $r$,

\begin{equation} \label{eq:tphi6}
y_{j}^{2} = \frac{r_{j}^{2}}{f(r_{j})}
\end{equation}
allows the metric function $f(r)$ to be fully reconstructed from the
plot of the endpoints, using a method of approximating the integral
similar to that in section \ref{sec:method1}.

\section{Analysing the two methods} \label{sec:eval}

Whilst the previous methods both in principle allow the bulk metric
to be fully recovered (from $f(r_{n})$ down to $f(0)$, with $r_{n}$
arbitrarily large), there are a number of factors to take into
account when performing the reconstruction in practice, in order to
obtain a balance between accuracy and computational effort.

Consider a setup where we divide the range of $y$ into $N + 1$ equal
segments, with $y_{0} = 0$ and $y_{N+1} = 1$, so that we have $N +
1$ steps in the process of recovering $f(r)$ (Note that the first
step in the iteration uses $y_{N}$ rather than $y_{N+1}$, as we want
the first minimum radius to be finite).

There are a number of issues with recovering $f(r)$ in this
scenario. We firstly observe that splitting $y$ up into equal
segments means that the minimum radii will not be equally separated;
there will be a greater number of low radius points (due to the
shape of $V_{eff}$). This initially appears helpful, as we expect
most of the bulk metric's deviation from pure AdS to be localised
close to the centre, and so a greater number of data points in this
region should improve our estimate of $f(r)$.

However, as both methods for recovering $f(r)$ involve approximating
integrals over $r$, accuracy is only maintained if the step size is
kept relatively small for all $r$; whilst starting with $y_{N}$ as
close to one as possible allows $f(r)$ to be recovered from as large
a radius as possible (as higher $y_{N}$ means higher $r_{N}$), this
also means the first few steps in $r$ will be unacceptably large.
This can be countered by ensuring the number of steps is
appropriately large, however, this will then result in a longer
computational time. For example, in the asymptotically AdS spacetime
with $f(r)$ given by

\begin{equation} \label{eq:fexample1}
f(r) = 1 + r^{2} - \frac{4 r^{2}}{(r^{2} + 1)(r^{2} + 8)}
\end{equation}
if use a step size in y of $0.0005$ and take the initial $y_{N}$ to
be $0.9995$, the corresponding minimum radius $r_{N} \approx 31.5$.
The minimum radius corresponding to $y_{N-1} = 0.999$ is only
$r_{N-1} \approx 22.3$ however, and such a large jump in the radius
causes problems with using our approximation of the form of
\eqref{eq:trnminus1finalm1}. Whilst the original Laurent expansion
of \eqref{eq:trnminus1dm1} is still fairly accurate over this
distance, our approximation of the gradient by

\begin{equation} \label{eq:fgradm2}
f'(r_{n-i}) = \frac{f(r_{n-i+1}) - f(r_{n-i})}{r_{n-i+1} - r_{n-i}}
\end{equation}
is not. For the example above, the actual value of $f'(r)$ at $r =
22.3$ is 44.5, whereas the value given by \eqref{eq:fgradm2} is
53.8; a large discrepancy.

We can still split the $y_{n}$ linearly (in order to keep the
majority of the points at low radii), however, we choose the initial
$y_{N}$ to be be slightly lower. This enables our first step in the
radius to be kept small.\footnote{This can be checked in practice by
calculating $r_{N}$ and $r_{N - 1}$ for a given choice of initial
$y_{N}$; if the difference $r_{N} - r_{N - 1}$ is too great, either
lower $y_{N}$ or increase $N$ until it's acceptable.} In our example
of \eqref{eq:fexample1}, if we choose $y_{N} = 0.9985$ and keep the
step size at $0.0005$, we find that $r_{N} - r_{N-1} \approx 2.5$
which whilst still quite large, is much more acceptable than the
value of $9.2$ we had previously. Table \ref{table0} shows how this
discrepancy in the gradient affects the extraction of $f(r)$ by the
method of section \ref{sec:method1} for the example of
\eqref{eq:fexample1}; it is worth noting at this stage that even
with the lower choice of starting $y$ the recovered estimate for
$f(r)$ is still quite poor with a step size of $0.0005$. This is
important because although we could lower $y_{N}$ even further, we
must keep our initial radius reasonably large; if we choose $y_{N}$
to be too low, we risk having $r_{N}$ being too small, such that our
assumption that the spacetime there is approximately that of pure
AdS is no longer valid.\footnote{This poses an interesting question
as to how one determines what a ``reasonable'' initial radius is for
an unknown spacetime. If our $r_{N}$ was too low, such that the
metric was not approximately pure AdS at this point, how would this
be apparent from the extracted estimate of $f(r)$? A simple solution
is to check the behaviour of the estimate at $r$ close to $r_{N}$;
if the function does not continue as $r^{2} + 1$ (i.e. pure AdS)
down to say $r = r_{N}/2$, then the chosen initial radius was too
small. It is also worth noting that in a physical situation, one
would naturally expect ordinary matter to remain within a radius of
order of $R$ from the centre, due to the confining AdS potential.}
We can keep our choice of $y_{N}$ high by reducing the step size (as
in the lower half of Table \ref{table0}), but at the expense of
additional computational time (For more details on the numerical
extraction see the examples in the next section).

\begin{table}
\begin{center}
\begin{tabular}{l l l l l}
\hline Initial y & Step size & $\alpha$ (4) & $\beta$ (1)& $\gamma$ (8)\\
\hline
0.9995 & 0.0005 & 0.778 & 0.971 & 0.971 \\
0.9985 & 0.0005 & 1.39 & 1.52 & 1.52 \\
\hline
0.9995 & 0.00005 & 2.37 & 2.24 & 2.24 \\
0.9985 & 0.00005 & 3.42 & 1.13 & 6.18 \\
\hline
\end{tabular}
\end{center}
\caption{Comparing the accuracy of using two different values for
$y_{N}$ to recover the bulk metric information. The actual values
for $\alpha$, $\beta$ and $\gamma$, which correspond to the three
numerical factors in \eqref{eq:fexample1} (and are properly defined
in \eqref{eq:ffit1}), are given in brackets. More details on how
these estimates were generated is given in section 5.}\label{table0}
\end{table}

There is, however, another problem with numerically reconstructing
$f(r)$ using the approximation of the gradient by
\eqref{eq:fgradm2}. If we examine the form of the term containing
$f'(r)$, namely the $A_{n - i}$ in \eqref{eq:trnminusim1}, whose
equivalent in method II is:

\begin{equation} \label{eq:anminusi2}
A_{n-i} = \frac{4}{3} \frac{(r_{n-i+1} - r_{n-i})}{f(r_{n-i})}
\sqrt{\frac{2}{r_{n-i}} - \frac{f'(r_{n-i})}{f(r_{n-i})}}
\end{equation}
we note that the $f'(r)$ appears with a negative sign inside the
square root. As our approximation of $f'(r)$ by \eqref{eq:fgradm2}
is in general an overestimation of the gradient, as for large $r$,
$f(r) \propto r^{2} + 1$, when solving for $r_{n-i}$ and
$f(r_{n-i})$ the term under the square root can become negative,
leading to the estimates being imaginary. This can occur even if the
overestimation is small.

There are a number of possible ways in which to resolve this
problem. We can firstly consider a different approximation of the
gradient to the one given in \eqref{eq:fgradm2}; the over-estimation
can be avoided by using an expression such as

\begin{equation} \label{eq:fgradm333}
f'(r_{n-i}) = \frac{1}{2} \Big( \frac{f(r_{n-i+1}) -
f(r_{n-i})}{r_{n-i+1} - r_{n-i}} + \frac{f(r_{n-i}) -
f(r_{n-i-1})}{r_{n-i} - r_{n-i-1}} \Big)
\end{equation}
which takes the average of the two nearest linear tangents to the
curve. Inserting this expression into either of our iterative
methods for recovering $f(r)$, however, immediately raises a new
problem from a computational point of view; we now cannot recover a
value for $r_{n-i}$ and $f(r_{n-i})$ at each step, as the $n-i$ term
now also depends on the subsequent term, $n-i-1$, as well as the
previous. We now have to determine all the equations for the various
steps and solve them together at the end to recover
$f(r)$.\footnote{Note that the final gradient $f'(r_{0})$ will have
to be approximated using an expression of the form of
\eqref{eq:fgradm2}} This is a much more complicated operation than
numerically solving at each step, and leads to a considerable
increase in the computational effort required to reconstruct the
metric function.

Another approach is to take a higher order series expansion around
the minimum radius in our approximation of the integral, however we
then need an expression for the second (or higher) derivative of
$f(r)$, which leads to an escalation of the problems mentioned
above; either the overestimation caused by only using approximations
involving larger $r$ terms upsets the recovery of $r_{n-i}$ and
$f(r_{n-i})$, or we need to wait and solve all the equations at the
end of all the steps, again greatly increasing the computational
effort required.

The gradient of $f(r)$ appears in the integral of
\eqref{eq:trnminusim1} because of the Laurent expansion we have used
in its approximation; in method II, however, we can use a different
approximation which avoids this.

\begin{figure}
\begin{center}
  \includegraphics[width=0.8\textwidth]{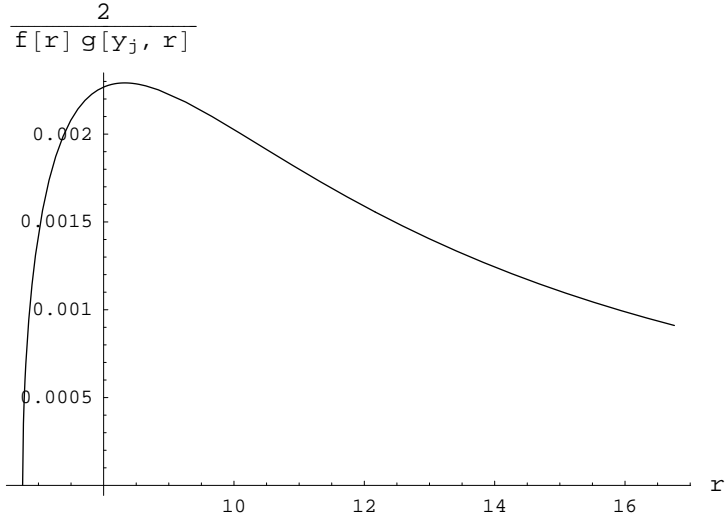}\\
\end{center}
\caption{A plot showing an example curve (the integrand of
\eqref{eq:tphi5}) to be approximated in method II. The leftmost part
of the curve is initially a vertical parabola.}\label{para1}
\end{figure}

Fig.\ref{para1} shows a plot of a typical curve we are trying to
approximate. The crucial difference between this type of curve and
that of Fig.\ref{traps1} is the behaviour at the minimum radius,
which is due to the positioning of the $g(y,r)$ term in equations
\eqref{eq:trnminusim1} and \eqref{eq:tphi5}. That $g(y,r)$ appears
in the denominator of \eqref{eq:tphi5} rather than the numerator
results in the curve of Fig.\ref{para1} having an infinite gradient
at the minimum radius. We can then describe the leftmost part of the
curve as following a vertical parabolic path. Thus instead of using
a Laurent approximation at this point, we can make a geometric one
and say that:

\begin{equation} \label{eq:geometric1}
A_{n-i} = \frac{4}{3} \frac{(r_{n-i+1} - r_{n-i})}{f(r_{n-i+1})}
\sqrt{1 - \frac{y_{n-i}^{2}}{y_{n-i+1}^{2}}}
\end{equation}
where we have used that the area under a parabola is $2/3$ the width
multiplied by the height. The gradient of the curve will always be
infinite at the minimum radius, and as the other terms ($B_{n-i}$
and $C_{n-i}$) do not depend on $f'(r)$, we avoid the need for an
approximation to the gradient of the form of \eqref{eq:fgradm2}.
Importantly, we also avoid any increase in the computational effort
required to recover $f(r)$. It is also worth noting that there is no
similar geometric argument for such an approximation to be applied
in the original method of section \ref{sec:method1}. For similar
step size this modification leads to a considerably more accurate
estimation of $f(r)$ from the endpoints, as we shall see in the next
section, where we use both methods to reconstruct $f(r)$ in two
different cases.

\section{Examples}

\begin{figure}
\begin{center}
  \includegraphics[width=0.9\textwidth]{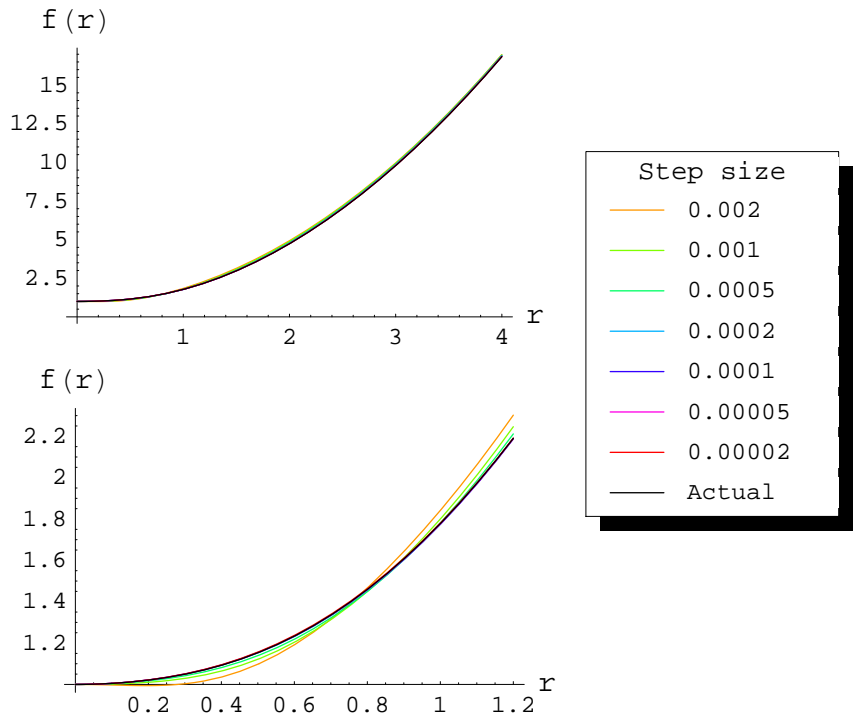}\\
\end{center}
\caption{Plots showing the various estimates for $f(r)$ given in
Table \ref{table1} (i.e. produced using method I), compared to the
actual metric function $f(r)$ from \eqref{eq:fexample1}. Whilst all
the estimates seem good fits over a large radius (top figure),
closer consideration of the curves highlights their differences
(bottom figure), which lead to the inaccurate values calculated for
$\alpha$, $\beta$ and $\gamma$ from the larger step size
estimates.}\label{examp1}
\end{figure}

Consider a small deviation from pure AdS, such as the spacetime with
metric function $f(r)$ as in \eqref{eq:fexample1}. Choosing our
initial value of $y$ to be $y_{N} = 0.9985$ (for the reasons of the
previous section), we investigate the ability of the two methods to
extract the bulk information. Fig.\ref{examp1} shows our
approximations of $f(r)$ generated using method I for a range of
different step sizes.

For this first example, the top set of curves in Fig.\ref{examp1}
suggest that we are easily able to extract the bulk metric using any
of the step sizes, as they all appear to lie very close to the
actual function $f(r)$. However, this is somewhat misleading, as we
shall see. In order to more closely examine the accuracy of the
recovered bulk data, we use a non-linear fit of the form:

\begin{equation} \label{eq:ffit1}
f(r) = 1 + r^{2} - \frac{\alpha \, r^{2}}{(r^{2} + \beta)(r^{2} +
\gamma)}
\end{equation}
to obtain values for $\alpha$, $\beta$ and $\gamma$ for each
estimate, and the results are presented in Table \ref{table1}. As we
mentioned above, looking at the top plot of Fig.\ref{examp1}, all of
the approximations seem very close to the original function,
however, Table \ref{table1} shows this not to be the case. Until we
go to a very small step size, we are unable to accurately extract
what one might consider the important metric information; if the
modification of the spacetime caused by the extra term in
\eqref{eq:fexample1} was to correspond to some physical phenomenon,
we might expect its ``mass'' and ``extent'' to be related to the
quantities $\alpha$, and $\beta$ and $\gamma$ respectively. Whilst
the best fits we obtain are converging to the correct values as we
take the step size smaller, we are already calculating a significant
number of terms to generate these estimates.

\begin{table}
\begin{center}
\begin{tabular}{l l l l l}
\hline Initial y & Step size & $\alpha$ (4) & $\beta$ (1)& $\gamma$ (8)\\
\hline 0.9985 & 0.002 & 0.371 & 0.530 & 0.530 \\
0.9985 & 0.001 & 0.797 & 0.995 & 0.995 \\
0.9985 & 0.0005 & 1.39 & 1.52 & 1.52 \\
0.9985 & 0.0002 & 2.18 & 2.11 & 2.11 \\
0.9985 & 0.0001 & 2.74 & 1.47 & 3.92 \\
0.9985 & 0.00005 & 3.42 & 1.13 & 6.18 \\
0.9985 & 0.00002 & 3.80 & 1.11 & 7.20 \\
\hline
\end{tabular}
\end{center}
\caption{Best fit values (to 3 s.f.) for $\alpha$, $\beta$ and
$\gamma$ for data generated using method I, with the actual values
indicated in brackets. Only in the lower half of the table do the
estimates for the three unknowns really start to converge to the
correct values.}\label{table1}
\end{table}

\begin{table}
\begin{center}
\begin{tabular}{l l l l l}
\hline Initial y & Step size & $\alpha$ (4)& $\beta$ (1)& $\gamma$ (8)\\
\hline 0.9985 & 0.002 & 1.61 & 2.01 & 2.01 \\
0.9985 & 0.001 & 3.34 & 1.16 & 6.01 \\
0.9985 & 0.0005 & 3.92 & 1.01 & 7.77 \\
0.9985 & 0.0002 & 3.99 & 1.00 & 7.97 \\
0.9985 & 0.0001 & 4.00 & 1.00 & 8.00 \\
\hline
\end{tabular}
\end{center}
\caption{Best fit values (to 3 s.f.) for $\alpha$, $\beta$ and
$\gamma$ for data generated using method II, with the actual values
indicated in brackets. Using the alternative method, we are able to
pick out rough values for the three unknowns as early as with a step
size of $0.001$, and by a step size of $0.0001$ the estimates have
converged to the correct values (to 3 s.f.).}\label{table2}
\end{table}

\begin{figure}
\begin{center}
  \includegraphics[width=0.9\textwidth]{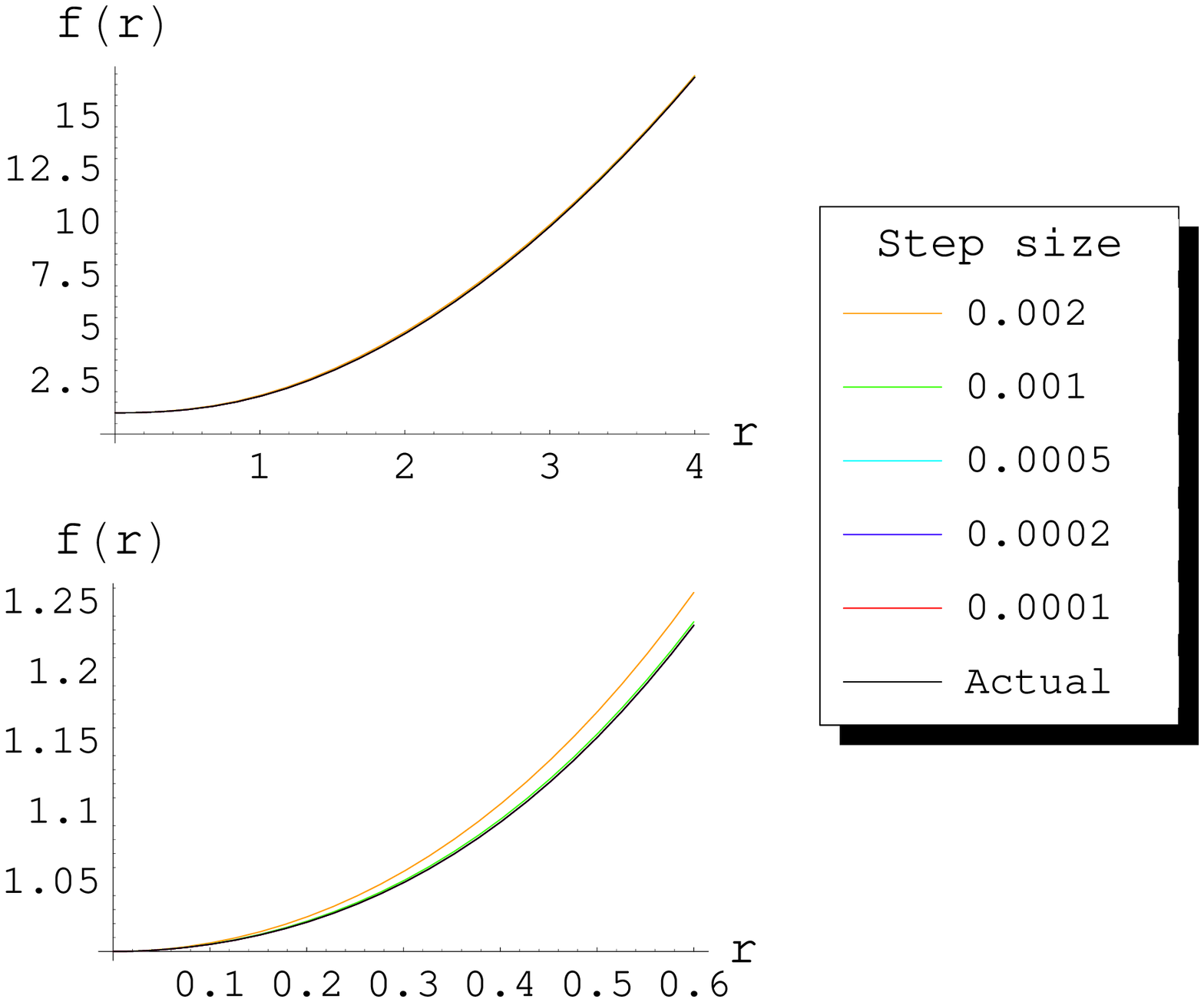}\\
\end{center}
\caption{Plots showing the various estimates for $f(r)$ given in
Table \ref{table2}, compared to the actual metric function $f(r)$
from \eqref{eq:fexample1}. Using method II for producing the
estimates has resulted in much closer fits, both over a large radius
(top figure), and at smaller scales (bottom figure). This is
reflected in the highly accurate values of $\alpha$, $\beta$ and
$\gamma$ determined from the all but the largest step size
estimates.}\label{examp11}
\end{figure}

If we compare this to the fits we obtain using the data calculated
via method II, with the modification to the $A_{n-i}$ terms given in
\eqref{eq:geometric1}, we see a notable difference in the results.
The estimates again all look good when plotted with the actual
function $f(r)$, as we see in Fig.\ref{examp11}, however, as the
bottom plot shows, this accuracy is now maintained at smaller
scales. Table \ref{table2} shows the new values for $\alpha$,
$\beta$ and $\gamma$ for each step size, obtained using the same
non-linear fit, \eqref{eq:ffit1}, as before.

Comparing these estimates with those in Table \ref{table1}, we see
that the recovered values for $\alpha$, $\beta$ and $\gamma$ are far
more accurate using this alternative method, for each choice of step
size. We now have a very good fit to the actual values of 4,1 and 8
in \eqref{eq:fexample1} obtained using a relatively low number of
steps, and thus relatively quickly. Indeed, we see that to obtain
the same degree of accuracy using method I we need to use step sizes
which are more than twenty five times smaller, which (assuming a
standard computational time per step) would therefore take at least
twenty five times longer to compute. Thus using the modified second
method offers a significant improvement over the original.

\begin{figure}
\begin{center}
  \includegraphics[width=0.9\textwidth]{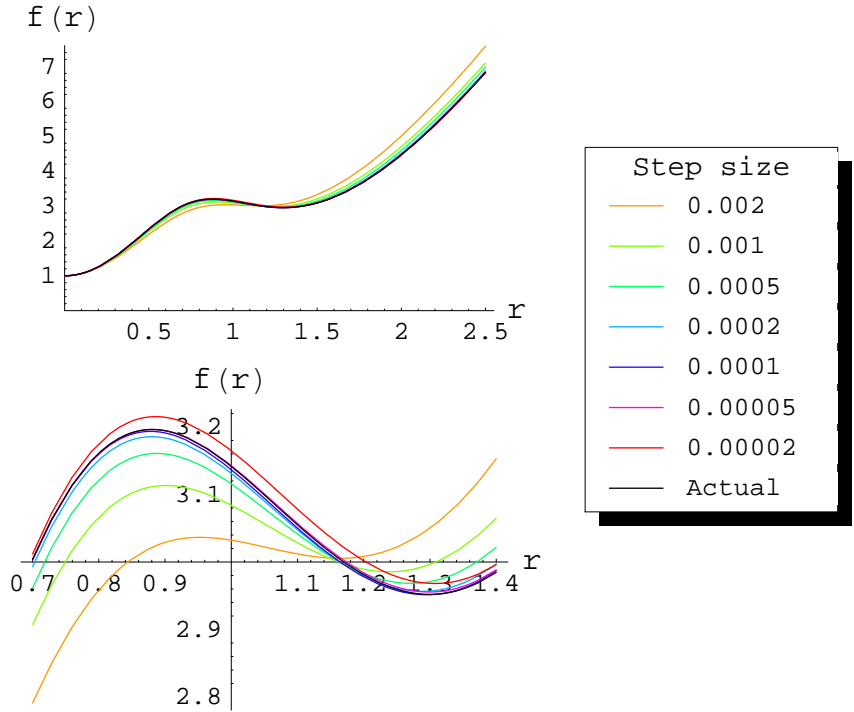}\\
\end{center}
\caption{Plots showing the various estimates for $f(r)$ given in
Table \ref{table3} (i.e. produced using method I), compared to the
actual metric function $f(r)$ from \eqref{eq:fexample2}. The
convergence of the estimates to the actual curve can be seen in both
plots, and whilst the values obtained for $\chi$, $\eta$ and
$\lambda$ are reasonably good for all step sizes, only upon using
the smallest step size do we start to pick out the $\alpha$, $\beta$
and $\gamma$ values. This is why, in the lower figure, the red curve
representing the final estimate appears to be a worse fit to the
actual curve than some of the others; the range of $r$ shown
highlights the accuracy of the $\chi$, $\eta$ and $\lambda$ values
rather than the $\alpha$, $\beta$ and $\gamma$ ones.}\label{examp2}
\end{figure}

\begin{figure}
\begin{center}
  \includegraphics[width=0.9\textwidth]{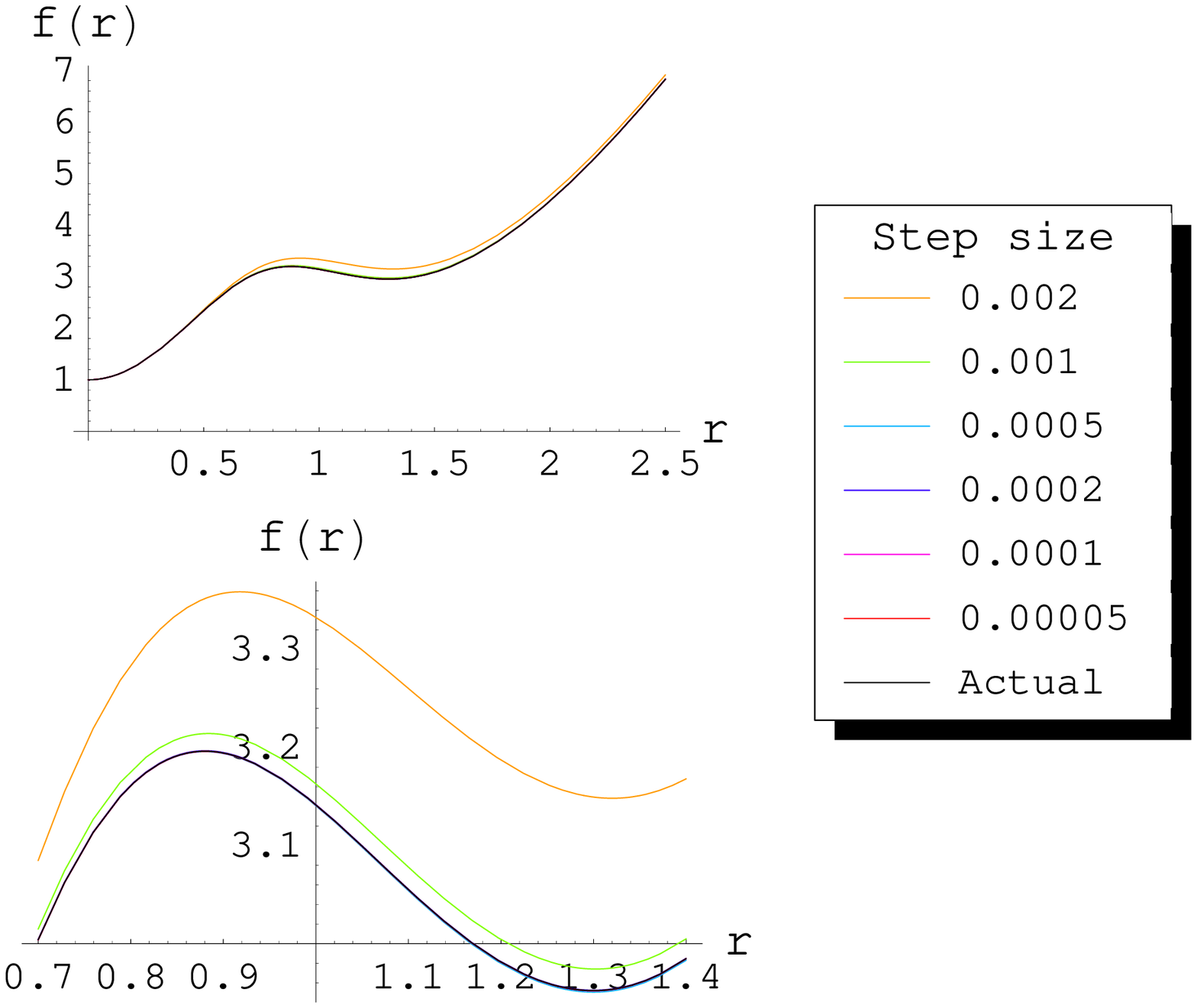}\\
\end{center}
\caption{Plots showing the various estimates for $f(r)$ given in
Table \ref{table4} (i.e. produced using method II), compared to the
actual metric function $f(r)$ from \eqref{eq:fexample2}. Only the
estimates generated using the largest two step sizes are visibly
distinguishable from the actual curve, as the the values of
$\alpha$, $\beta$, $\gamma$, $\chi$, $\eta$ and $\lambda$ are all
quickly picked out using this alternative method.}\label{examp22}
\end{figure}

Can we use the methods to recover a more complicated $f(r)$? If we
consider a spacetime similar to that in \eqref{eq:fexample1}, but
which is further modified at low $r$ by an extra term:

\begin{equation} \label{eq:fexample2}
f(r) = 1 + r^{2} - \frac{4 r^{2}}{(r^{2} + 1)(r^{2} + 8)} + \frac{3
r \sin(2 r)}{r^{4} + 1}
\end{equation}
we can use both methods to generate estimates for $f(r)$, and using
a non-linear fit of the form of:

\begin{equation} \label{eq:ffit2}
f(r) = 1 + r^{2} - \frac{\alpha \, r^{2}}{(r^{2} + \beta)(r^{2} +
\gamma)}+ \frac{\chi r \sin(\eta r)}{r^{4} + \lambda}
\end{equation}
we can judge the accuracy of the estimate against
\eqref{eq:fexample2}. Using the same starting value for $y$ and the
same step sizes as before, we obtain the results shown in tables
\ref{table3} and \ref{table4}, and figures \ref{examp2} and
\ref{examp22}.

\begin{table}
\begin{center}
\begin{tabular}{l l l l l l l l}
\hline Initial y & Step size & $\alpha$ (4)& $\beta$ (1)& $\gamma$ (8)& $\chi$ (3)& $\eta$ (2)& $\lambda$ (1)\\
\hline 0.9985 & 0.002 & -1380 & 133 & 133 & 2.60 & 2.20 & 1.20 \\
0.9985 & 0.001 & 0.0824 & 0.426 & 0.426 & 2.69 & 2.12 & 1.04 \\
0.9985 & 0.0005 & 0.790 & 0.880 & 0.880 & 2.96 & 2.03 & 0.981 \\
0.9985 & 0.0002 & 1.69 & 1.25 & 1.25 & 3.06 & 1.95 & 0.940 \\
0.9985 & 0.0001 & 2.23 & 1.62 & 1.62 & 3.05 & 1.94 & 0.946 \\
0.9985 & 0.00005 & 2.87 & 2.39 & 2.39 & 3.00 & 1.97 & 0.985 \\
0.9985 & 0.00002 & 3.68 & 1.31 & 6.50 & 3.09 & 2.00 & 1.04 \\
\hline
\end{tabular}
\end{center}
\caption{Best fit values (to 3 s.f.) for $\alpha$, $\beta$,
$\gamma$, $\chi$, $\eta$ and $\lambda$ for data generated using
method I, with the actual values indicated in brackets. In this more
complicated modification to AdS, the presence of the sine term masks
the finer structure and prevents the original method from converging
towards the correct values of $\beta$ and $\gamma$ until our
smallest choice of step size.}\label{table3}
\end{table}

\begin{table}
\begin{center}
\begin{tabular}{l l l l l l l l}
\hline Initial y & Step size & $\alpha$ (4)& $\beta$ (1)& $\gamma$ (8)& $\chi$ (3)& $\eta$ (2)& $\lambda$ (1)\\
\hline 0.9985 & 0.002 & 1.17 & 1.46 & 1.46 & 3.47 & 1.94 & 1.12 \\
0.9985 & 0.001 & 2.90 & 2.52 & 2.52 & 3.03 & 1.97 & 0.998 \\
0.9985 & 0.0005 & 3.85 & 1.06 & 7.38 & 3.00 & 2.00 & 1.00 \\
0.9985 & 0.0002 & 3.98 & 1.01 & 7.93 & 3.00 & 2.00 & 1.00 \\
0.9985 & 0.0001 & 4.00 & 1.00 & 7.99 & 3.00 & 2.00 & 1.00 \\
0.9985 & 0.00005 & 4.00 & 1.00 & 8.00 & 3.00 & 2.00 & 1.00 \\
\hline
\end{tabular}
\end{center}
\caption{Best fit values (to 3 s.f.) for $\alpha$, $\beta$,
$\gamma$, $\chi$, $\eta$ and $\lambda$ for data generated using
method II, with the actual values indicated in brackets. Once again,
the alternative method proves much more adept at accurately
estimating the unknowns, and converges to the correct values almost
immediately.}\label{table4}
\end{table}

Once again we see that the second method produces a much better
estimate of $f(r)$ than the first; the difference in the accuracy is
again significant. What is also noticeable in this slightly more
complicated example is that the presence of a more prominent feature
in the metric function (in this case the small oscillation from the
sine term) can obscure the finer details of the spacetime. This is
most obvious in the results from using method I (see Table
\ref{table3}, where we can clearly see that whilst the numerical
factors corresponding to the sine term ($\chi$, $\eta$ and
$\lambda$) are easily determined even at a large step size, it then
requires a smaller step size to pick out the $\alpha$, $\beta$ and
$\gamma$ values than it did in the first example.

Whilst the second example has shown that the methods can be used for
$f(r)$ which have a significant deviation from pure AdS near the
centre, there remain problems in applying the methods in certain
circumstances, which are examined in the next section.

\section{Limitations}

There are limitations in recovering $f(r)$ all the way down to $r =
0$ in the cases where the metric is non-singular but does not give a
monotonic effective potential for the null geodesics, and also when
there is a curvature singularity at $r = 0$.

\subsection{Metrics with a non-monotonic effective potential}

\begin{figure}
\begin{center}
  \includegraphics[width=0.8\textwidth]{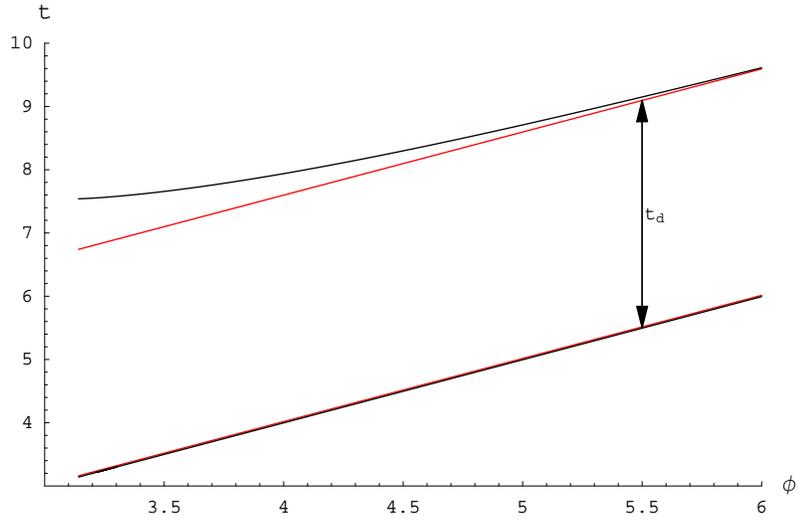}\\
\end{center}
\caption{Plot of the endpoints of null geodesics which all begin
from the same point on the boundary and pass through an AdS-like
space time with non-monotonic $V_{eff}$. The red lines indicate the
limits to which the black curves tend as $L \rightarrow
L_{crit}$}\label{Mend2}
\end{figure}

There is a problem with the method for reconstructing $f(r)$ if the
metric is such that for some critical value of the angular momentum,
the null geodesics can go into an unstable orbit (assuming fixed
energy). This is indicated on the endpoint plot by the $\phi$
coordinate heading off to infinity, and the presence of a jump in
the time taken by the geodesics with $L < L_{crit}$ compared with
those with $ L > L_{crit}$, see Fig.\ref{Mend2} for an example.

\begin{figure}
\begin{center}
  \includegraphics[width=0.8\textwidth]{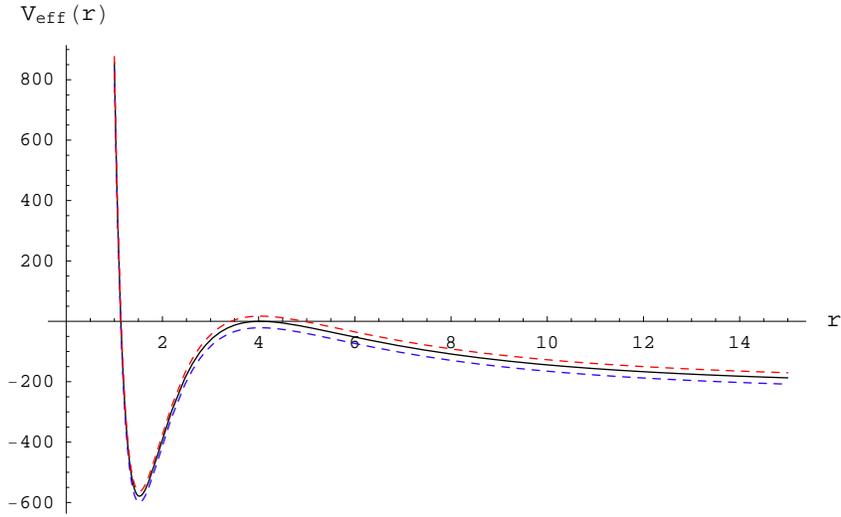}\\
\end{center}
\caption{Plot of the effective potential for three null geodesics:
one with the critical angular momentum, $L_{crit}$ (the solid black
curve), one with $L < L_{crit}$ (the lowest curve, dashed blue), and
one with $L > L_{crit}$ (top curve, dashed red).}\label{Mv1}
\end{figure}

The reason for these unstable orbits is because the effective
potential for the metric isn't monotonic; the orbits occur when the
greatest radius ($r_{p}$ say) for which $dV/dr = 0$ corresponds with
that for which $V = 0$, see Fig.\ref{Mv1}. This non-monotonicity
also explains the time delay for those geodesics with angular
momentum lower than $L_{crit}$, indicated by the gap between the
asymptotes in Fig.\ref{Mend2}: those with slightly lower $L$ have an
effective potential of the form shown as the dashed blue curve in
Fig.\ref{Mv1}, and travel down to minimum radius $r_{p-1}$, whereas
those with $L \ge L_{crit}$ have minimum radius greater than or
equal to $r_{p}$, indicated by the dashed red curve. The extra
(non-negligible) distance $r_{p} - r_{p-1}$ travelled by the
geodesics results in the time delay seen in the endpoints.

We thus have the situation where we have no geodesics with minimum
radius $r_{q}$, where $r_{p-1} < r_{q} < r_{p}$, and this is where
our iterative methods break down. Although we can begin by using one
of the methods from earlier to recover $f(r)$ down to $r = r_{p}$,
which corresponds to using the endpoints in the ``lower branch'' of
Fig.\ref{Mend2}, we cannot continue past this point. Using method I,
say, the final term of the iteration will be:

\begin{equation} \label{eq:num6d}
t_{p} = 2 \int_{r_{p}}^{\infty} \frac{1}{f(r) \sqrt{1 - y_{p}^{2}
\frac{f(r)}{r^{2}}}} \, \mathrm{d} r \approx A_{p} + B_{p} + C_{p}
\end{equation}
with $A_{p}$, $B_{p}$ and $C_{p}$ defined as in section
\ref{sec:method1}. As the next term would involve an integration
over the range $r_{p-1}$ to $r_{p}$, which cannot necessarily be
taken to be small, it cannot be well approximated by the methods we
have used previously.

The gap between the upper and lower branches of Fig.\ref{Mend2}
tends to a constant time delay, $t_{d}$, as $\phi \rightarrow
\infty$, and it is this which corresponds to the integral we are
unable to well approximate:

\begin{equation} \label{eq:timedelay1}
t_{d} = 2 \int_{r_{p-1}}^{r_{p}} \frac{1}{f(r) \sqrt{1 - y_{p-1}^{2}
\frac{f(r)}{r^{2}}}} \, \mathrm{d} r
\end{equation}
where we note that in the limit $\phi \rightarrow \infty$ we have
that $y_{p-1} = y_{p}$.

We are left with the situation where we do not have enough
information to carry on recovering $f(r)$ past the rightmost maximum
in the effective potential; there is no approximation we could
attempt to make without already knowing more about the form of
$V_{eff}$. Although we have the extra endpoint data (i.e. the top
branch of \ref{Mend2}), we are unable to use it to extract f(r).

Having discovered this apparent limitation in the extraction of the
bulk metric, we should ask ourselves how often we would expect to
find metrics which give non-monotonic effective potentials for the
null geodesics. In other words, how often do we find spacetimes
which allow stable orbits of light rays? Whilst it is very easy to
construct metrics by hand which allow this, in physical systems
circular orbits of light rays are a rarity. Indeed, recent work by
Hubeny \textit{et al.}, \cite{hubenynew}, indicates that for metrics
corresponding to a gas of radiation in AdS, it is impossible for
circular null geodesics to occur.

Admittedly our restrictive case of metrics of the form of
\eqref{eq:AdSmetric} are unrealistic, as any metric of this form
necessarily has opposite sign pressure and energy density components
of the stress tensor, and as our methods currently do not work in
the more general metrics, \eqref{eq:newAdSmetric}, this is merely an
observation.

One scenario in which null geodesics can go into circular orbits is
around black holes, and we now conclude the current section by
considering such an example.

\subsection{Metrics containing a singularity}

We ask what information can be retrieved if the metric contains a
singularity at its centre, for example Schwarzschild-AdS. We again
cannot recover the entirety of $f(r)$, as the geodesics which pass
behind the horizon will not return out to the boundary. Thus whilst
in the non-monotonic case we had extra information (the top branch
of Fig.\ref{Mend2}) available but no way to get past the jump in the
minimum radii of the geodesics, here we only have a reduced spectrum
of endpoints from which to recover $f(r)$.

The endpoints we do obtain are those with angular momentum above
some critical value, and so we can use the above iterative method to
recover $f(r)$ down to a certain critical radius (which will be
greater than the horizon radius). However, any null geodesics which
pass behind the horizon will always terminate at the singularity,
and so we are unable to obtain any further information using our
iterative method.\footnote{As mentioned in the introduction, there
are numerous papers devoted to the extraction of behind the horizon
information, and we do not concern ourselves with this here.}

For an example we consider five dimensional Schwarzschild-AdS, in
which the metric function $f(r)$ is given by:

\begin{equation} \label{eq:f3}
f(r) = 1 + \frac{r^{2}}{R^{2}}- \frac{r_{h}^{2}}{r^{2}}
\left(\frac{r_{h}^{2}}{R^{2}} +1\right)
\end{equation}
where $R$ is again the AdS radius, and $r_{h}$ is the horizon
radius. We set $R = 1$ as usual, and note that the effective
potential for null geodesics is given by:

\begin{equation} \label{eq:genSAdSveff3}
V_{eff} = \left(1 + \frac{r^{2}}{R^{2}}- \frac{r_{h}^{2}}{r^{2}}
\left(\frac{r_{h}^{2}}{R^{2}} +1\right)\right) \frac{L^{2}}{r^{2}} -
E^{2}
\end{equation}

For a null geodesic to avoid hitting the singularity at $r = 0$, we
require $V_{eff} = 0$ for some $r > r_{h}$. The lowest $L$ for which
this will happen will be when the gradient of $V_{eff}$ is also
zero. Differentiating \eqref{eq:genSAdSveff3} with respect to $r$,
and setting equal to zero gives

\begin{equation} \label{eq:genSAdSveffd}
\frac{2 L^{2}}{r^{5}} \left(2\left(r_{h}^{2} + r_{h}^{4}\right) -
r^{2}\right) = 0
\end{equation}
which is equal to zero when $r = \sqrt{2} \sqrt{r_{h}^{2} +
r_{h}^{4}}$. Thus for the case where $r_{h} = 1$ say, the minimum
$r$ which can be probed by the null geodesics is $r = 2$, and hence
we can obtain information about the metric function $f(r)$ from $r =
\infty$ down to $r = 2$.

\section{Discussion}

We have seen that the endpoints of null geodesics can be used to
recover information about the bulk in asymptotically AdS spacetimes.
By varying the ratio of angular momentum to energy of the geodesics
we are able to systematically probe to different radii, and this,
combined with the appropriate AdS boundary conditions, allows us to
reconstruct the metric. What is perhaps quite surprising is that
this local, geometric information about the spacetime can be
completely recovered in certain scenarios; in a metric of the form
of \eqref{eq:AdSmetric}, as long as the effective potential for the
null geodesics is monotonic, we are able to completely recover the
function $f(r)$. Whilst this is of course due to the fact that we
have severely restricted our spacetime in such a scenario, to ensure
that the amount of information we require is contained within the
bi-local spectrum of endpoints, it is still impressive that the
recovery process itself is so simple. Its simplicity lies in the
remarkable nature of our expression for the gradient of the
endpoints of null geodesics \eqref{eq:dtdphib}; to have that the
relative angular momentum of the geodesic corresponds to the
endpoint in such an accessible fashion is astounding, and forms the
basis of our ability to reconstruct the bulk.

Focusing on the cases with a monotonic $V_{eff}$, we developed two
different methods for carrying out the recovery of the bulk
information, both of which involved iteratively reconstructing
$f(r)$ from large $r$ down to the centre. In the process of actually
using these methods to numerically extract the bulk information, we
made several refinements to increase their efficiency, the most
important of which involved our approximation of the gradient of
$f(r)$. In both our original methods, we were consistently
overestimating $f'(r)$, a problem that was significantly affecting
their accuracy, unless we kept the iterative step size very small.
Whilst we suggested a number of ways of resolving this problem, all
made the process of recovering $f(r)$ considerably more complex to
implement, thus requiring a great deal of extra computational
effort. Given that one of the main aims in this work was to produce
a practical method for the recovery of the metric information, these
resolutions were not acceptable. Rather than investigating further
methods of approximating $f'(r)$, we looked at the question of why
we needed an approximation to the gradient of $f(r)$ at all, as
$f'(r)$ does not appear anywhere in our original geodesic equations.

The solution lay in finding an alternative to the Laurent series we
were using to approximate the section of the geodesic path close to
its minimum radius. Whilst this had been the natural approximation
to use in our first method, there was a simpler one we could apply
in the second, which stemmed from a crucial difference in the type
of curve we were trying to approximate. In method I, the curve in
question is infinite at the minimum radius, because the $g(y,r)$
term (see \eqref{eq:brevity}) appears in the numerator of
\eqref{eq:trnminusim1}. In method II, on the other hand, the
$g(y,r)$ term appears in the denominator of the relevant equation,
\eqref{eq:tphi5}, which results in the curve being zero with an
infinite gradient. This allowed us to approximate the curve
geometrically by a vertical parabola, and thus the integral by the
corresponding parabolic area formula. This modification enabled us
to estimate $f(r)$ far more efficiently, as it removed any reference
to $f'(r)$. It also greatly improved the ability of the iterative
process to extract what one might consider the important metric
information: the numeric values relating to the ``mass'' and
``extent'' of whatever physical phenomenon was causing the deviation
from pure AdS.

We concluded the project by considering two alternative scenarios:
where the effective potential was non-monotonic, and where the
metric contained a singularity at the centre. In each case, we were
only able to recover $f(r)$ down to a certain critical radius,
corresponding to a local maximum of $V_{eff}$ (see fig.\ref{Mv1}).
We were unable to continue the process of recovering $f(r)$ any
further as, in the non-singular case, there is a non-negligible
``jump'' in the minimum radii of the geodesics, due to the shape of
the effective potential, which we cannot well approximate. Thus
although we have additional endpoint data, we are unable to use it
to extract more of the bulk information; we do, however, note that
in physical situations, which is where we would eventually hope to
apply these techniques, we do not often encounter such metrics (i.e.
spacetimes in which light rays can enter into circular orbits). For
those metrics containing a singularity, we do not have any further
endpoints with which to continue recovering $f(r)$, as any null
geodesic travelling to lower radii will pass behind the horizon and
terminate at the singularity.

A possible extension of this work is in considering less restrictive
forms of the metric, such as those where we have relaxed spherical
symmetry, or where the metric can evolve with time. Whilst our
previous attempt (in section 3.3) at modifying the iterative methods
to work in a more general spacetime failed due to the extra unknowns
we introduced, this does not mean every modification would be so
problematic. One can envisage, for example, a situation in which the
metric function fluctuates over time, in such a way that the
effective potential for the null geodesics remains monotonic
throughout. In this scenario, whilst the process might be more
complicated, we should still be able to extract the entire bulk
metric, including the time dependence, from the complete set of null
geodesic endpoints.

Another natural extension of this project would be to consider the
endpoints of spacelike geodesics, and examine whether they can be
used to recover more information about the bulk in the various
scenarios. The main issue with producing a method to iteratively
recover the metric information as we have done here with the null
geodesics is that our equation \eqref{eq:dtdphi} involving the
gradients of the endpoints is significantly more complicated. For
spacelike geodesics we have two independent parameters ($L$ and $E$)
to consider; simply using their ratio as we did in the null case is
no longer sufficient. Other works have examined this in more detail,
e.g. \cite{fest}, and they give encouragement that progress can be
made in this direction. The challenge then would be whether a
computationally practical method can also be found.

\section*{Acknowledgements}

I would like to thank Veronika Hubeny for insightful discussions and
useful feedback throughout this work, which was supported by an
EPSRC studentship grant.
\\
\section*{Appendix A: Null geodesic paths in AdS space}

For the null geodesic paths we have $\dot x^{2} = 0$, and we can
solve the following three equations numerically to provide the plots
of the geodesics:

\begin{equation} \label{eq:AdSnull1}
E = \left(1 + \frac{r^{2}}{R^{2}}\right) \dot t
\end{equation}
\begin{equation} \label{eq:AdSnull2}
L = r^{2} \dot \phi
\end{equation}
\begin{equation} \label{eq:AdSnull3}
- \left(1 + \frac{r^{2}}{R^{2}}\right) \dot t^{2} + \frac{\dot
r^{2}}{1 + \frac{r^{2}}{R^{2}}} + r^{2}\dot \phi^{2} = 0
\end{equation}

It is also useful to calculate the geodesics analytically. This can
be done most easily by rewriting the original equation for the
metric \eqref{eq:AdSmetric} in a new form. Rescaling the radial and
time coordinates $r \rightarrow R \tan(r)$ and $t \rightarrow R \,
t$ gives:

\begin{equation} \label{eq:AdSmetricrescaled}
ds^{2} = - R^2\left(\sec^{2}(r) dt^{2} + \sec^{2}(r) dr^{2} +
\tan^{2}(r) d\phi^{2}\right)
\end{equation}
where we have suppressed two of the angular coordinates as before.
This leads to the modified constraint equations:

\begin{equation} \label{eq:AdSnullmod1}
E = R^2 \sec^{2}(r) \dot t
\end{equation}
\begin{equation} \label{eq:AdSnullmod2}
L = R^2\tan^{2}(r) \dot \phi
\end{equation}
\begin{equation} \label{eq:AdSnullmod3}
- R^2\left(\sec^{2}(r) \dot t^{2} + \sec^{2}(r) \dot r^{2} +
\tan^{2}(r) \dot \phi^{2}\right) = 0
\end{equation}

Combining the three equations, by substituting
\eqref{eq:AdSnullmod1} and \eqref{eq:AdSnullmod2} into
\eqref{eq:AdSnullmod3} to eliminate $\dot t$ and $\dot \phi$ we
have:

\begin{equation} \label{eq:AdSnull4}
- \frac{E^{2}}{R^2\sec^{2}(r)} + R^{2} \sec^{2}(r) \dot r^{2} +
 \frac{L^{2}}{R^{2} \tan^{2}(r)} = 0
\end{equation}

Rearranging, and again using \eqref{eq:AdSnullmod1} to eliminate the
dependence on $\lambda$ gives:

\begin{equation} \label{eq:AdSnull5}
\sqrt{\frac{E^{2}}{\left(R^2\sec^{2}(r)\right)^{2}} -
\frac{L^{2}}{R^{4} \tan^{2}(r) \sec^{2}(r)}} = \dot r = \frac{d r}{d
t} \frac{d t}{d \lambda} = \frac{d r}{d t} \frac{E}{R^2\sec^{2}(r)}
\end{equation}

and so
\begin{equation} \label{eq:AdSnull6}
\frac{d r}{d t} = \sqrt{1 - \frac{L^{2}}{E^{2} \sin^{2}(r)}}
\end{equation}

Using the substitution $x = \frac{\cos(r)}{\sqrt{1 - L^{2}/E^{2}}}$
this equation becomes:

\begin{equation} \label{eq:AdSnull7}
\frac{d x}{d t} = - \sqrt{1 - x^{2}}
\end{equation}
which implies $x = - \sin (t + t_{0})$ and thus that

\begin{equation} \label{eq:AdSnull8}
\cos(r) = - \sqrt{1 - \frac{L^{2}}{E^{2}}} \sin (t + t_{0})
\end{equation}

We can also calculate the dependence of $\phi$ on $t$:

\begin{equation} \label{eq:AdSnull9}
\frac{L}{R^{2} \tan^{2} (r)} = \dot \phi = \frac{d \phi}{d t}
\frac{d t}{d \lambda} = \frac{d \phi}{d t} \frac{E}{R^{2}
\sec^{2}(r)}
\end{equation}
thus:

\begin{equation} \label{eq:AdSnull10}
\frac{d \phi}{d t} = \frac{L}{E \sin^{2}(r)}
\end{equation}
which can be combined with \eqref{eq:AdSnull8} to give:

\begin{equation} \label{eq:AdSnull11}
\tan(\phi + \phi_{0}) = \frac{L}{E} \tan (t + t_{0})
\end{equation}

Note that the dependence of the geodesics on the AdS radius of
curvature has been scaled out in \eqref{eq:AdSnull8} and
\eqref{eq:AdSnull11} by the choice of coordinates.

\section*{Appendix B: Cancelling the divergent term from the
Leibniz rule}

To make sure our expression for the derivative of $t$ with respect
to $y$ is finite, we have to combine the two terms from:

\begin{eqnarray} \label{eq:dtrappendix21}
\frac{d t(y)}{dy}&& \hspace{-0.5cm} = 2 \frac{d}{dy}
\left(\int_{r_{min}}^{\infty} \frac{g(y,r)}{f(r)} \, \mathrm{d} r
\right)
\\ \label{eq:dtrappendix22}
&& \hspace{-0.5cm} = 2 y \int_{r_{min}}^{\infty}
\frac{(g(y,r))^{3}}{r^{2}} \, \mathrm{d} r - \left(\frac{2 \,
g(y,r)}{f(r)}\right)\Big|_{r = r_{min}} \frac{d r_{min}}{dy}
\end{eqnarray}
to eliminate the divergence piece. We need an expression for $d
r_{min}/dy$, obtained using our expression for $y$ in terms of the
minimum radius, \eqref{eq:tphi6},

\begin{equation} \label{eq:appendix23}
\frac{d r_{min}}{dy} = 1 \Big/ \left( \frac{d y}{d r} \Big|_{r =
r_{min}} \right) = \frac{1}{\sqrt{f(r_{min})}}\left(1 -
\frac{r_{min} f'(r_{min})}{2 \, f(r_{min})}\right)
\end{equation}

which we then use to rewrite the second term of
\eqref{eq:dtrappendix22} as an integral from $r_{m}$ to $\infty$:

\begin{eqnarray*} \label{eq:dtrappendix24}
\left(\frac{2 g(y,r)}{f(r)}\right)\Big|_{r = r_{m}} \frac{d
r_{m}}{dy} && \hspace{-0.5cm} = \int_{r_{m}}^{\infty}
\frac{\left(f'(r) -
r f''(r)\right)\sqrt{f(r)} \, g(y,r)}{\left(f(r) - \frac{r}{2} f'(r)\right)^{2}} \, \mathrm{d} r \\
&& - \int_{r_{m}}^{\infty} \frac{f'(r) \, g(y,r)}{\left(f(r) -
\frac{r}{2} f'(r)\right) \sqrt{f(r)}} \, \mathrm{d} r
\\ && + \int_{r_{m}}^{\infty} \frac{2 \, y \, \left(g(y,r)\right)^{3}}{r^{2}} \, \mathrm{d} r
\end{eqnarray*}

We see that the third term above cancels precisely with the
divergent integral in \eqref{eq:dtrappendix22}, and after
rearranging we are left with

\begin{equation} \label{eq:dtrappendix25}
\frac{d t(y)}{dy} = \int_{r_{m}}^{\infty} \frac{r
\left(f'(r)\right)^{2} - 2 \, r f''(r) \, f(r) \, g(y,r)}{2
\left(f(r) - \frac{r}{2} f'(r)\right)^{2} \sqrt{f(r)}} \, \mathrm{d}
r
\end{equation}
which is finite.

\section*{Appendix C: Attempting to recover the bulk metric in the
more general spacetime}

In the more general case of a spacetime described by equation
\eqref{eq:newAdSmetric}, we have two unknown functions $f(r)$ and
$h(r)$ to recover, along with $r$ itself, and hence we need three
linearly independent equations. We appear to have them in equations
\eqref{eq:trc}, \eqref{eq:phirc} and \eqref{eq:tphi6}.

Following the method of section \ref{sec:method1}, we take $r$ large
enough so that the spacetime is approximately pure AdS, and solve

\begin{equation} \label{eq:appendix31}
y_{n}^{2} = \frac{r_{n}^{2}}{f(r_{n})} = \frac{r_{n}^{2}}{r_{n}^{2}
+ 1}
\end{equation}
to give $r_{n}$, and thus also $f(r_{n}) = 1/h(r_{n}) = r_{n}^{2} +
1$.

We now consider a geodesic with slightly lower $y$ and follow the
method as before, except that we now use the expressions for $t$ and
$\phi$ given in \eqref{eq:trc} and \eqref{eq:phirc}. Beginning with
the expression for $t$, we split it up as in \eqref{eq:trnminus1bm1}
and evaluate the second integral as in \eqref{eq:trnminus1ccm1}.
This leaves

\begin{equation} \label{eq:appendix3a}
t_{n-1} \approx 2 \int_{r_{n-1}}^{r_{n}}
\frac{1}{\frac{f(r)}{\sqrt{h(r)}} \sqrt{\frac{1}{f(r)} -
\frac{y_{n-1}^{2}}{r^{2}}}} \, \mathrm{d} r + \pi - 2 \arctan{\left(
\sqrt{\left(1 - y_{n-1}^{2}\right)r_{n}^{2} - y_{n-1}^{2}}\,
\right)}
\end{equation}
where we can approximate the first integral by its lowest order
Laurent expansion around the minimum radius. This gives

\begin{equation} \label{eq:appendix3b}
t_{n-1} \approx \frac{4 \sqrt{r_{n} - r_{n-1}}
\sqrt{h(r_{n-1})}}{\sqrt{\frac{2 f(r_{n-1})}{r_{n-1}} -
f'(r_{n-1})}} + \pi - 2 \arctan{\left( \sqrt{\left(1 -
y_{n-1}^{2}\right)r_{n}^{2} - y_{n-1}^{2}}\, \right)}
\end{equation}

Similarly for the expression for $\phi$ we obtain

\begin{equation} \label{eq:appendix3c}
\phi_{n-1} \approx \frac{4 \sqrt{r_{n} - r_{n-1}}
\sqrt{h(r_{n-1})}}{y_{n-1} \sqrt{\frac{2 f(r_{n-1})}{r_{n-1}} -
f'(r_{n-1})}} + 2 \arctan{\left(\frac{y_{n-1}}{ \sqrt{\left(1 -
y_{n-1}^{2}\right)r_{n}^{2} - y_{n-1}^{2}}}\, \right)}
\end{equation}
which initially appears to be linearly independent of
\eqref{eq:appendix3b}, as although the first term in
\eqref{eq:appendix3c} is simply the first term in
\eqref{eq:appendix3b} divided by $y_{n-1}$, the two second terms
appear different. Even rewriting the second and third terms of
\eqref{eq:appendix3b} using

\begin{equation} \label{eq:appendix3d}
\frac{\pi}{2} - \arctan{\left( \sqrt{\left(1 -
y_{n-1}^{2}\right)r_{n}^{2} - y_{n-1}^{2}}\, \right)} =
\arctan{\left(\frac{1}{ \sqrt{\left(1 - y_{n-1}^{2}\right)r_{n}^{2}
- y_{n-1}^{2}}}\, \right)}
\end{equation}
does not linearly relate to the second term of
\eqref{eq:appendix3c}, as there is the extra $y_{n-1}$ factor inside
the $\arctan$.

The degeneracy occurs because we are attempting to use these two
equations, along with $y_{n-1}^{2} = r_{n-1}^{2}/f(r_{n-1})$, to
recover $r_{n-1}$, $f(r_{n-1})$ and $h(r_{n-1})$; we already have
values for the other variables. Thus the only term which is of
importance in recovering the unknowns is the first, and the apparent
independence of the two equations has arisen from our approximation
of the second term. Hence equations \eqref{eq:appendix3b} and
\eqref{eq:appendix3c} effectively both reduce to:

\begin{equation} \label{eq:appendix3e}
\frac{\sqrt{r_{n} - r_{n-1}} \sqrt{h(r_{n-1})}}{\sqrt{\frac{2
f(r_{n-1})}{r_{n-1}} - f'(r_{n-1})}} = constant
\end{equation}
We are thus left with only two equations to determine the three
unknowns, and so the method of extracting the metric functions
immediately breaks down.\footnote{Note that using a higher order
Laurent expansion in \eqref{eq:appendix3a} does not affect the
degeneracy of the equations.}

\end{document}